\documentclass[nologo,11pt,a4paper]{ETHpaper}
\usepackage{graphicx, amsmath, amssymb,color,wasysym}
\usepackage{bm}
\usepackage[square,numbers,sort&compress]{natbib}

\usepackage{longtable}
\usepackage{subfig}

\begin{document}

\newcommand{\mean}[1]{\left\langle #1 \right\rangle} 
\newcommand{\abs}[1]{\left| #1 \right|} 
\newcommand{\ul}[1]{\underline{#1}}
\renewcommand{\epsilon}{\varepsilon} 
\newcommand{\eps}{\varepsilon} 
\renewcommand*{\=}{{\kern0.1em=\kern0.1em}}
\renewcommand*{\-}{{\kern0.1em-\kern0.1em}} 
\newcommand*{\+}{{\kern0.1em+\kern0.1em}}

\newcommand{\RA}{\Rightarrow}
\newcommand{\bbox}[1]{\mbox{\boldmath $#1$}}

\title{International crop trade networks: The impact of shocks and cascades}

\titlealternative{International crop trade: Impact of shocks and cascades}
  
\author{Rebekka Burkholz, Frank Schweitzer}
\authoralternative{R. Burkholz, F. Schweitzer}

\address{Chair of Systems Design, ETH Zurich, Weinbergstrasse 58, 8092 Zurich, Switzerland}

\reference{(Submitted for publication: 17 January 2019)}

\www{\url{http://www.sg.ethz.ch}}

\makeframing
\maketitle

\begin{abstract}
  Analyzing available FAO data from 176 countries over 21 years, we observe an increase of complexity in the international trade of maize, rice, soy, and wheat.
A larger number of countries play a role as producers or intermediaries, either for trade or food processing.
  In consequence, we find that the trade networks become more prone to failure cascades caused by exogenous shocks.
  In our model, countries compensate for demand deficits by imposing export restrictions.
  To capture these, we construct higher-order trade dependency networks for the different crops and years.
  These networks reveal hidden dependencies between countries and allow to discuss policy implications.  

  \end{abstract}
\date{\today}

\section{Introduction}

The production and trade of food involves almost all countries in the world, this way forming a global network of dependencies.
This network is reconstructed and analyzed in our paper.
It reflects \emph{direct} import and export relations between countries and further serves as a basis to estimate how shocks of food production in one country impact other countries in an \emph{indirect} manner.
We focus on the international trade network of staple food, in particular  maize, rice, soy and wheat, as the most important sources of calories for human consumption.
Their amount traded internationally has vastly increased over the past two decades \cite{FAO2016}, for several reasons, in particular
a growing world population \cite{popGrowth}, increasing meat and feed consumption linked to economic growth \cite{IncomeMeat}, or demand for biofuels \cite{biofuel}.
Because of globalization, countries can specialize in the production of the food they have the appropriate resources for.
But a larger number of countries can also benefit from food trade and from adding value to food products by means of additional processing.
In consequence, if production or trade of this staple food is hampered, for instance because of natural catastrophes or political conflicts, many more countries are affected either directly or indirectly.

Our aim is to (i) reveal these direct and indirect dependencies using data from 176 countries over 22 years and (ii) to model the impact of different shock scenarios on the international food trade.
This is challenging because the roles of countries cannot be easily reduced to \emph{producers}, \emph{importers} and \emph{exporters}.
Some countries produce a given crop mainly for export (e.g. Brazil produces soy), whereas other countries obviously rely on the import of their staple food (e.g. Saudi Arabia imports substantial shares of its rice and wheat).
But many countries are important as intermediaries, either because of their role as traders (e.g. the Netherlands) or because they produce intermediate or final products from these staple foods (e.g. Italy).
Hence, even countries that produce a given staple food can appear as importers, while countries that do not produce a given staple food, can appear as exporters. 

This situation is tidily related to the advantages and disadvantages of globalized food trade.
On the one hand, involving more countries allows for specialization and more complex value chains.
Furthermore, global markets facilitate risk diversification \citep{Burkholz2015} to better mitigate supply shocks, e.g. harvest losses.
On the other hand, countries become more exposed to shocks through these global markets.
Such shocks can include, e.g., production losses due to weather anomalies, droughts, or pests.
Other disadvantages include larger costs to society and to climate because of global transportation. 
Further, more transshipment points increase food loss and facilitate the spreading of pests \cite{nopsa2015}.

While the resulting fragility of the global food system is already acknowledged \cite{Puma2015}, the propagation of shocks in international trade networks is not fully understood and their indirect consequences are not quantified. 
That's why, in the second part of our paper, we model how shocks of different sizes in one country impact
the availability of maize, rice, soy, or wheat in other parts of the world.
Specifically, the shock of a given country reduces its production or supply of a given crop.
If this results in an unmet demand, this country can reduce this trade flow by imposing export restrictions.
Such restrictions might motivate affected countries to do the same, which can trigger \emph{cascades of export restrictions}. 
Termed as multiplier effect, such cascades can also be observed empirically and are found to influence food prices \cite{Headey2011,exportban}.
The relevance of cascades was already discussed \cite{exportban,Puma2015}.
Yet, they are, to our knowledge, not yet implemented in stress tests that analyze the resilience of international grain trade \cite{MaizeNetworkAna,Puma2015,popGrowth} with one exception: \cite{grainCascades} considers an aggregated grain trade network and more complicated dynamics that assume that the cascade process actually models the time evolution of the network formation.
This, however, conflicts with having only information about aggregated trade between countries over one year.
Instead, we restrict ourselves to reveal indirect trade flows. 

In our model, we assume that the rules of global food trade do not change over time, to focus on effective trade flow dependencies between countries and the resulting cascades of export restrictions.
A similar model has been developed recently \cite{seafood} to study the vulnerability of global seafood trade with respect to shocks that are proportional to a shocked country's seafood exports.
Different to that model, we focus on the trade of maize, rice, soy, and wheat and consider shock scenarios that depend on the production and demand of countries.
This acknowledges the nature of most shocks possible and allows to study a country's exposure to local shocks in comparison to cascades that started in far distant countries.

Our model is data-driven, that is, it uses the available data to (i) reconstruct the international food trade network, which then is used to (ii) evaluate the global impact of different shock scenarios on the real network (which is different for every crop and for every year).
The resulting cascades of export restrictions are, in our case, not captured by the annual data and therefore difficult to assess.
Here, our model comes into play, as it provides insights into the unobserved indirect effects. 
We recall that the shock of a main producer of a given crop has not only consequences for the food supply of the own population or the population in other countries, it also impacts trade and food processing in intermediary countries.
To capture this impact, we propose a new method, the construction of a \emph{higher-order trade dependence network}.
While a first-order network only captures the impact of shocks on direct export partners, higher-order networks consider all indirect effects resulting from cascades of export restrictions. 
Our visualizations of the higher-order trade dependency network can be compared with other  studies about international food trade. 
For instance, the international rice and wheat trade networks from 1992-2009 are discussed by \cite{Puma2015} where 
also their vulnerability to export restrictions (without considering cascade effects) is analyzed.
Further, the authors of \cite{MaizeNetworkAna} identify the main actors in the international trade of maize and the trade structure from 2000 to 2009, while \cite{Diaz-bonilla2000} focus on clusters. 
Also related is the analysis of caloric and monetary trade flows \cite{MacDonald01032015} aggregating different food types and the development of a dynamic flux model to measure the countries' vulnerability to food contaminations \cite{Ercsey-Ravasz2012}.

\section{Data analysis and network construction}

\label{sec:data-analys-netw}

\subsection{Available data on the country level}
\label{sec:available-data-sets}

Food imported into a country can either be consumed by the population or further exported, either directly or after value is added, e.g. bread is produced from flour. 
The available data provided by the Food and Agricultural Organization of the United Nations \cite{dataTrade} at a resolution of \emph{one year} only gives total numbers about food production, import and export with respect to different countries.  
For our analysis, we consider data for $N=176$ countries over a period of 22 years, from 1992-2013.
This period is particularly interesting because, after the dissolution of the Soviet Union, from 1992 onwards geographic territories have been rather stable and an on-going globalization has shaped also the international food trade. 

\begin{figure}[htbp]
 \centering
\includegraphics[width=0.49\textwidth]{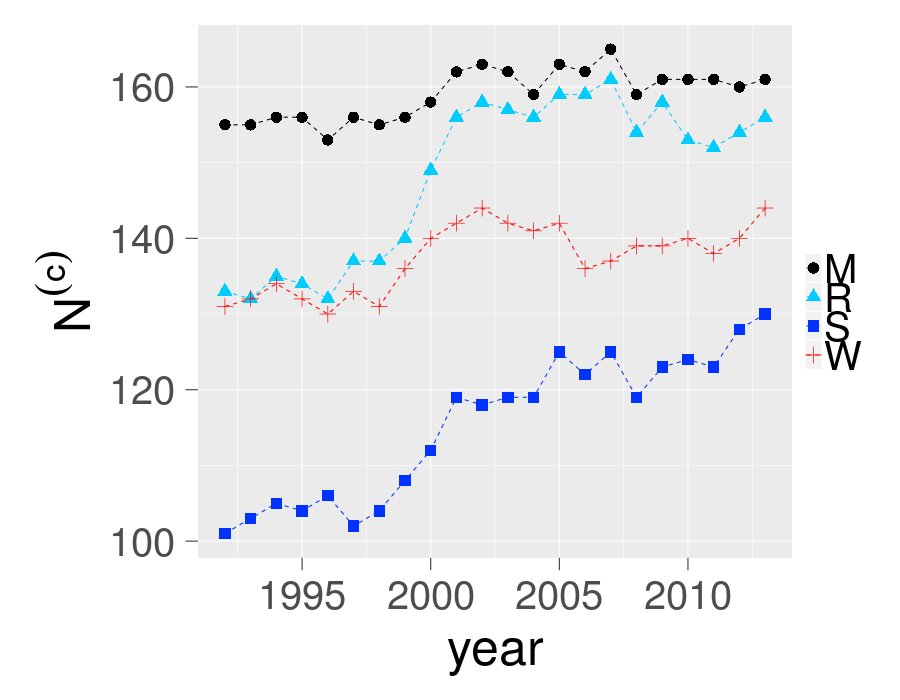}
\caption{Number of countries $N^{(c)}(y)$ that engage in trade or production of staple food $c\in{M,R,S,W}$ in a year $y$. $M$: maize, $R$: rice, $S$: soy, $W$: wheat.}
\label{fig:N}
\end{figure}
We consider four different crops, \emph{maize}, \emph{rice}, \emph{wheat} and \emph{soy}, because these are the main internationally traded crops and denote them with the index $c\in{M,R,S,W}$. 
$N^{(c)}(y)$ is the number of countries that engage in trade or production of crop $c$ in year $y$.
It is plotted in Fig.~\ref{fig:N} over time and tend to increase for all crops over the years.
However, since 2001/2002, $N^{(c)}(y)$ seems to stagnate for maize, rice, and wheat. 

\begin{figure}[htbp]
 \centering
\subfloat{(a)}{\includegraphics[width=0.45\textwidth]{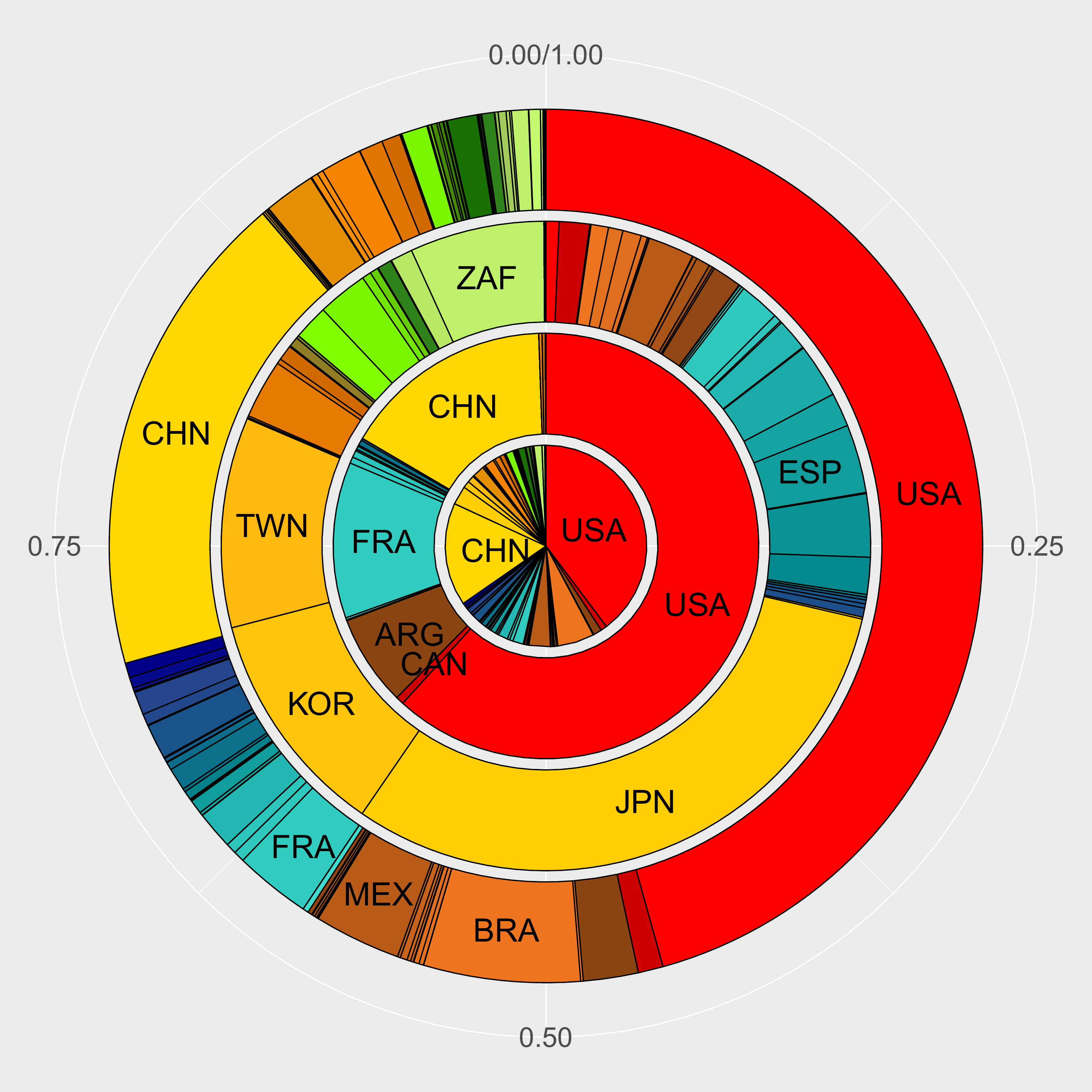}}\hfill
\subfloat{(b)}{\includegraphics[width=0.45\textwidth]{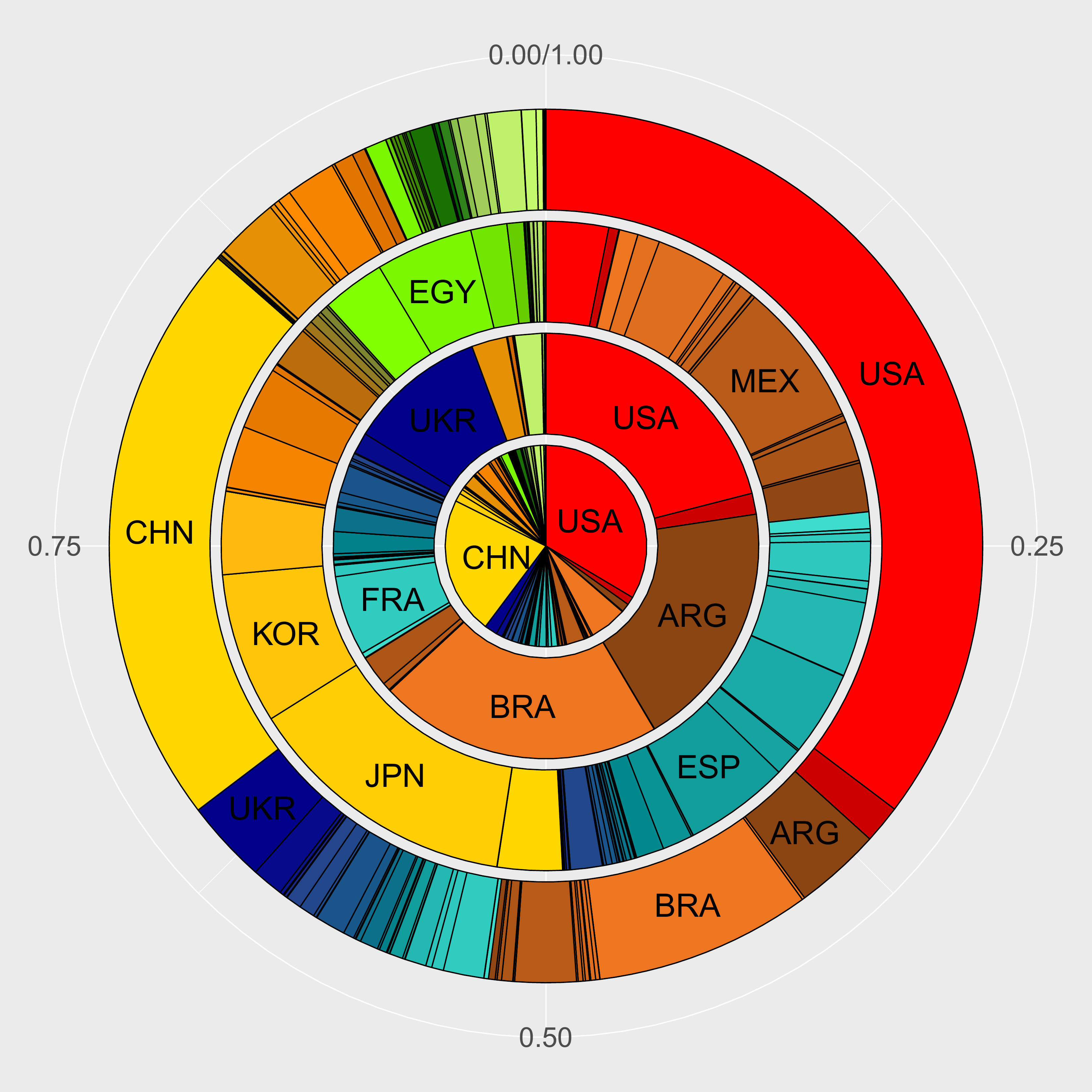}}
\caption{Fractions of maize \emph{production} $\mathrm{prod}_{i}^{(M)}(y)$ (outer circle), \emph{import} $\mathrm{imp}_{i}^{(M)}(y)$ (second outer circle), \emph{export} $\mathrm{exp}_{i}^{(M)}(y)$ (second inner circle) and \emph{demand} $\mathrm{dem}_{i}^{(M)}(y)$ (inner circle) per country in $y$=1992 (left) and $y$=2013 (right).
     Each figure should be read as the superposition of four separate pie charts. This allows a direct comparison of the respective quantities. Different colors indicate countries according to the world map shown in Fig.~\ref{fig:data}.
     Abbreviations follow the name convention given in Table~\ref{table:ctryList}. 
 The corresponding Figures for rice, soy and wheat are shown in Figures \ref{fig:pier}, \ref{fig:pies}, \ref{fig:piew} (Appendix).
   }
\label{fig:pieMaize}
\end{figure}
Our data set contains information about the annual \emph{production}, $\mathrm{prod}_{i}^{(c)}(y)$, of countries $i=1,...,N$ with respect to a given crop $c$, their exports, $\mathrm{exp}_{i}^{(c)}(y)$, and their imports, $\mathrm{imp}_{i}^{(c)}(y)$,  measured in tons.
From this, we can already calculate a country's demand for a given crop in a given year as:
\begin{align}
  \mathrm{dem}_{i}^{(c)}(y) = \mathrm{prod}_{i}^{(c)}(y) + \mathrm{imp}_{i}^{(c)}(y) - \mathrm{exp}_{i}^{(c)}(y) 
  \label{eq:2}
\end{align}
These numbers change over time and vastly differ across countries as Fig.~\ref{fig:pieMaize} shows.
For instance, the combined harvest of only the five biggest producers in 2013 amounts to ca. 89\% of the global soy, 79\% of the rice, 71\% of the maize, and 52\% of the wheat production.
Interestingly, as Fig.~\ref{fig:pieMaize} demonstrates, most countries are producers, importers and exporters of the same crop at the same time.
This already points to the complexity of worldwide food trade, because production shocks in a given country involve almost every other country via import and export.

\begin{figure}[htbp]
 \centering
\subfloat{(a)}{\includegraphics[width=0.45\textwidth]{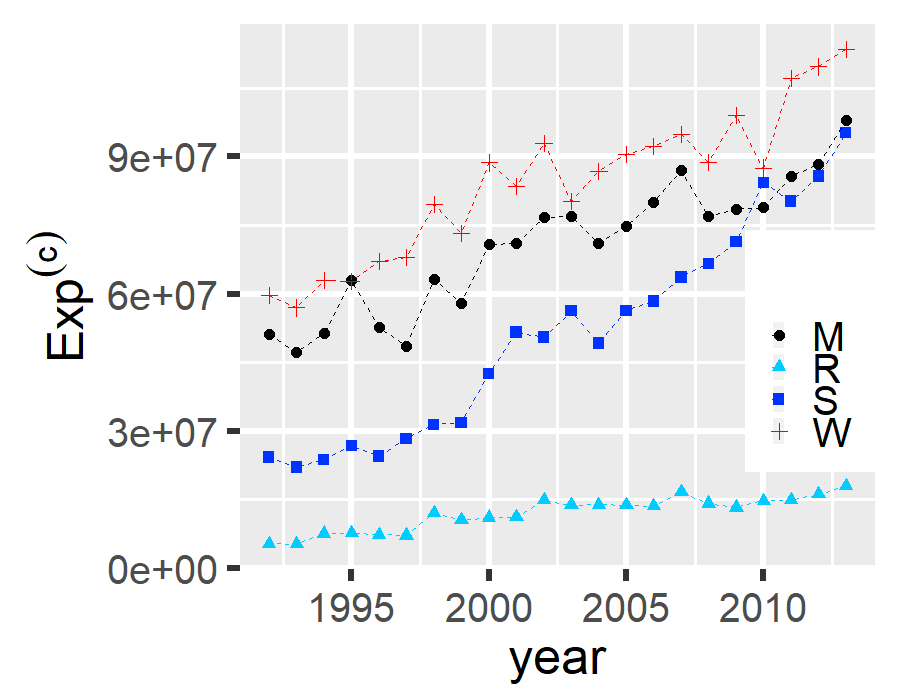}}
\hfill
 \subfloat{(b)}{\includegraphics[width=0.45\textwidth]{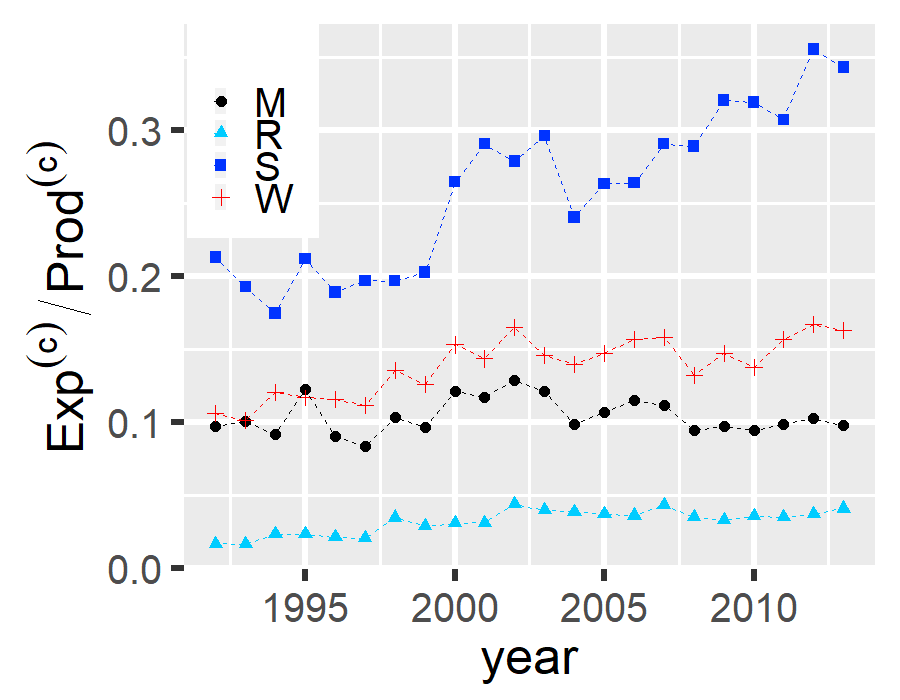}}
\caption{(a) Global exports $\mathrm{Exp}^{(c)}(y)$ in tons over time. (b) Global exports as a fraction of the total production, $\mathrm{Exp}^{(c)}(y)/\mathrm{Prod}^{(c)}(y)$, over time.}
 \label{fig:exp}
\end{figure}
Eventually, we can obtain the global exports, $\mathrm{Exp}^{(c)}(y)$, and the global production, $\mathrm{Prod}^{(c)}(y)$, as:
\begin{align}
  \label{eq:3}
 \mathrm{Exp}^{(c)}(y)=\sum_{i=1}^{N^{(c)}(y)} \mathrm{exp}_{i}^{(c)}(y) \;; \quad  \mathrm{Prod}^{(c)}(y)=\sum_{i=1}^{N^{(c)}(y)} \mathrm{prod}_{i}^{(c)}(y)
\end{align}
$\mathrm{Exp}^{(c)}(y)$ is plotted in Fig.~\ref{fig:exp}(a).
While the respective quantities steadily increase, it is more interesting to compare them with the annual global production,  $\mathrm{Prod}^{(c)}(y)$, of a given crop in the same year.
Fig.~\ref{fig:exp}(b) shows that total exports keep up to, or increase even faster, than the global production.
This fact should be valued against the observation in Fig.~\ref{fig:N} that the number of countries involved in production or trade of maize and rice is almost constant after the year 2000. 
Especially soy is traded internationally to a large extent, although the least number of countries participate in trade or production. Accordingly, soy trade is characterized by very high trade volumes.

\subsection{Constructing the trade networks}
\label{sec:constr-trade-netw}

In the following we construct from the available data the trade networks with respect to the different crops and the different years. 
Each country is represented by a node $i$ in a network $G^{(c)}(y)=\left(V^{(c)}(y), W^{(c)}(y)\right)$.
The set of all nodes is denoted by $V^{(c)}(y)$ with $N^{(c)}(y)$ elements. 
In total, we consider $N = 176$ countries. 
However, not all engage in trade or harvest crops in every year.
So, usually $N^{(c)}(y) < 176$.

\begin{figure}[htbp]
\centering
\includegraphics[width=\linewidth]{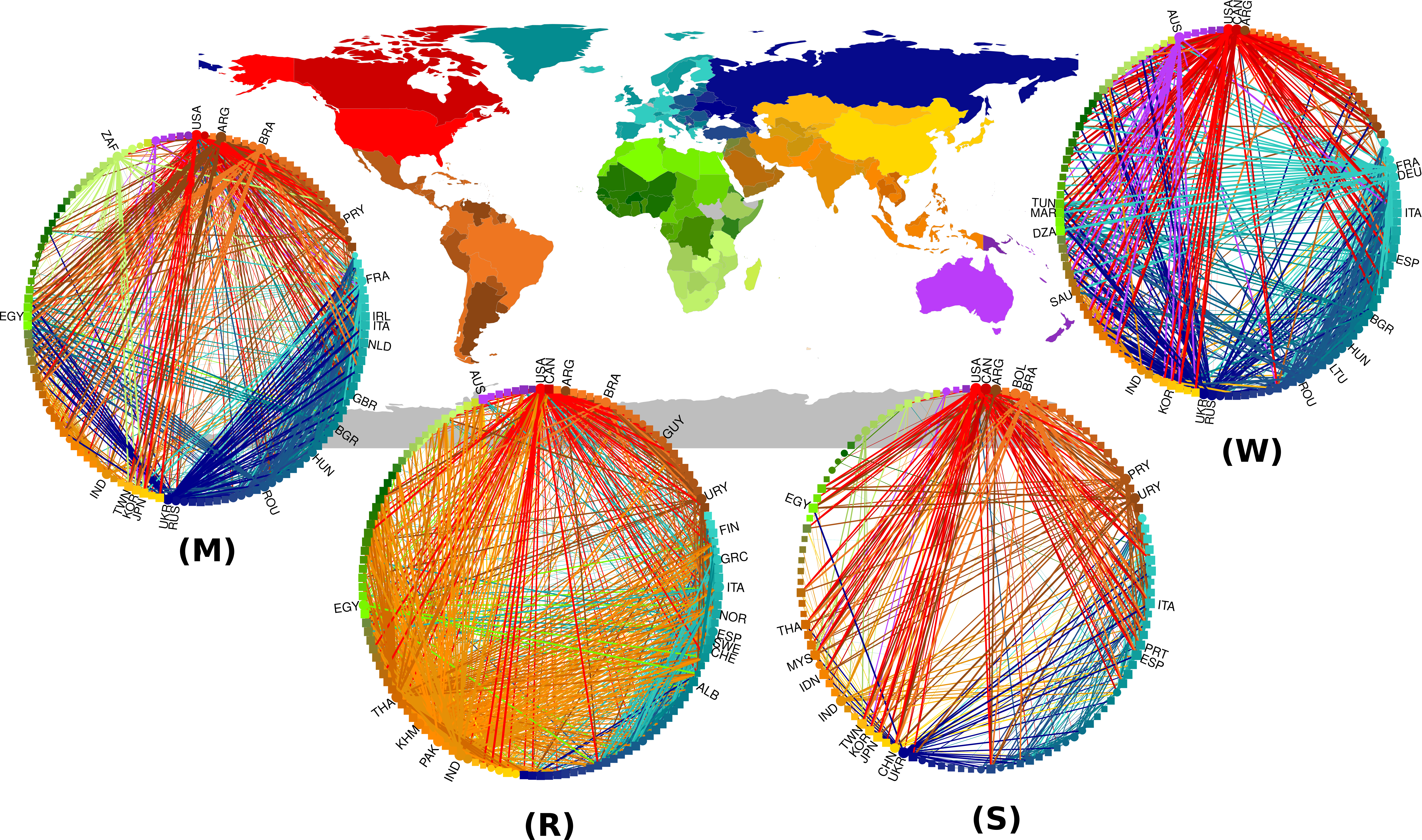}
\caption{International trade networks in 2013 for maize (M), rice (R), soy (S), and wheat (W).
  Each \emph{node} is colored according to the world map. The color of a \emph{link} $(i,j)$ corresponds to the exporting country $i$, with a link weight proportional to a logarithmic transformation of the export quantity: $\log\left(1 + w_{ij}\right)$. Links with larger weights are plotted on top of smaller ones. Square node shapes indicate that the respective country is a net importer, while circles refer to net exporters. The node size is proportional to a log transformation of their net imports or net exports. The twenty biggest nodes have their ISO-3 country code assigned (see Table~\ref{table:ctryList}). Isolated nodes (i.e. without connections) are omitted in a network plot.}\label{fig:data}
\end{figure}
Exports of crop $c$ from country $i$ to $j$ are represented by directed and weighted links,  $w^{(c)}_{ij}(y)\geq 0$. 
The set of all weighted links is denoted by $ W^{(c)}(y)$. 
On the basis of the weights $w_{ij}^{(c)}(y)$, we can express the total exports and imports of a given country $i$ as: 
\begin{align}
  \mathrm{exp}_{i}^{(c)}(y) = \sum^N_{j=1} w_{ij}^{(c)}(y) \;;\quad \mathrm{imp}_{i}^{(c)}(y) = \sum^N_{j=1} w_{ji}^{(c)}(y)
 \label{eq:1}
\end{align}
Their difference $\Delta_{i}^{(c)}(y)=\mathrm{exp}_{i}^{(c)}(y)-\mathrm{imp}_{i}^{(c)}(y)$ is used in Fig.~\ref{fig:data} to indicate net importers and net exporters. 

The trade networks $G^{(c)}(2013)=\left(V^{(c)}(2013), W^{(c)}(2013)\right)$ for the four different crops are visualized in Fig.~\ref{fig:data}.
We observe that the soy trade network has the lowest number of links.
However, the single trade volumes are comparatively large and, compared to the other three crops, the highest fraction of the total production is traded internationally (ca. 34\%).
In contrast, the rice trade network has the highest number of links, while its total trade volume sums up only to ca. 4\% of the total rice production, which is the smallest observed fraction.

We have also studied how the global trade networks of maize, rice, soy and wheat have evolved between 1992 and 2013. 
The plots of the empirical networks are shown in Figures \ref{fig:wMnet}\ref{fig:wRnet}\ref{fig:wSnet}\ref{fig:wWnet} in the Appendix.

\subsection{Change of network properties}
\label{sec:change-glob-netw}

\begin{figure}[htbp]
 \centering
 \subfloat{(a)}{\includegraphics[width=0.45\textwidth]{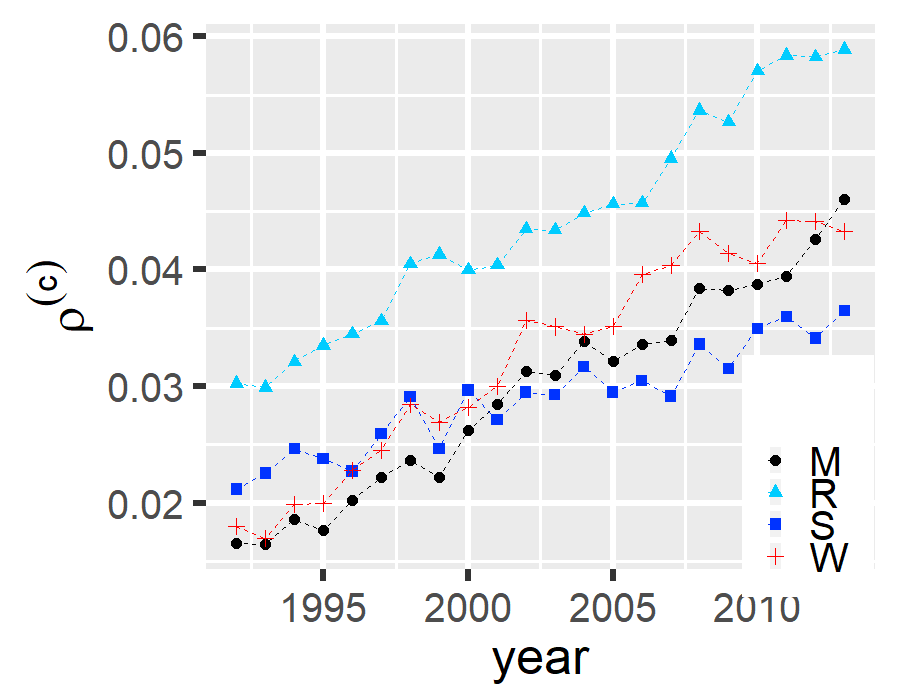}}
 \hfill
   \subfloat{(b)}{\includegraphics[width=0.45\textwidth]{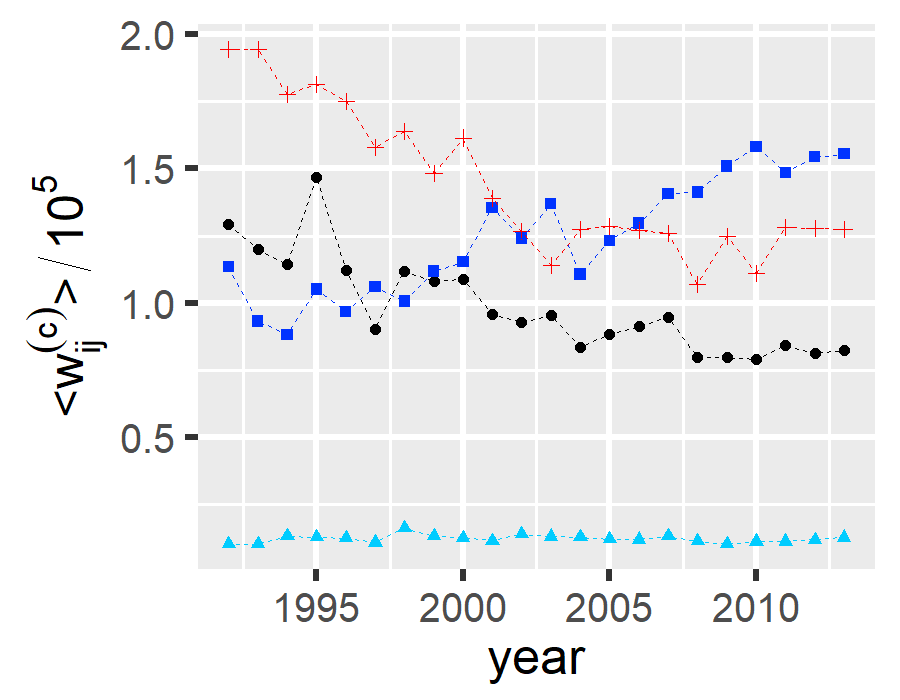}}
 \caption{Network density. (a) Link density: Fraction of links in comparison to fully connected network. (b) Average positive trade volume.}
\label{fig:netwDens}
 \end{figure}
The trade relationships between countries evolve over time, as illustrated in the following. 
Fig. \ref{fig:netwDens}~(a) depicts the change of the link density $\rho^{(c)}(y)=L^{(c)}(y)/\left(N^{(c)}(y)\left(N^{(c)}(y)-1\right)\right)$, where $L^{(c)}(y)$ denotes the number of all trade links in a network in year $y$.
The normalization is with respect to a fully connected network with $N(N-1)$ directed links.
As shown, $\rho^{(c)}(y)$  clearly increases over time, but not always at the same growth rate as the global exports shown in Fig.~\ref{fig:exp}.

\begin{figure}[htbp]
 \centering
 \includegraphics[width=0.49\textwidth]{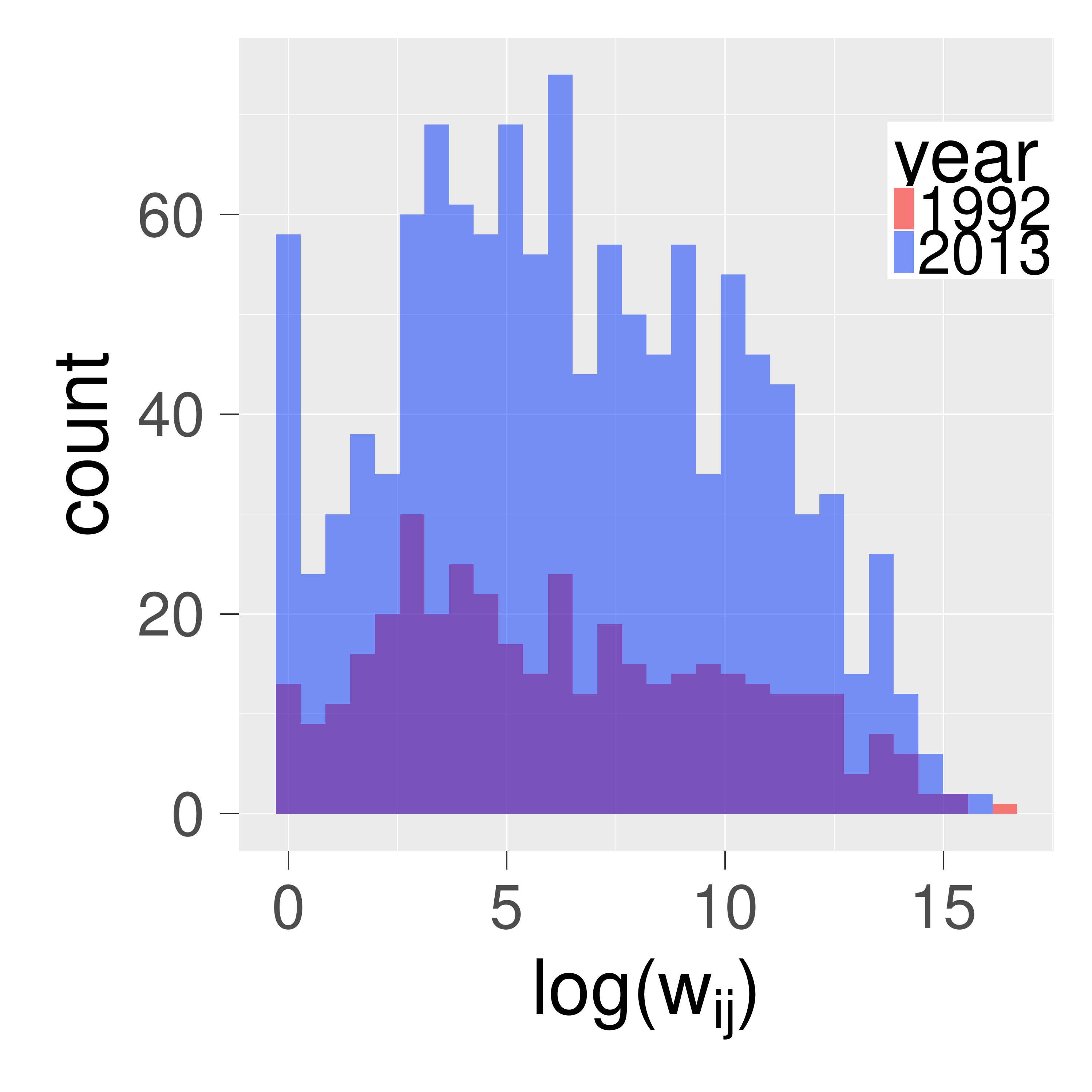}
 \caption{Histogram of the logarithm of the positive trade volumes in the years 1992 (red or purple if behind the blue) and 2013 (blue) for maize. The corresponding Figures \ref{fig:weights-soy} (a,b,c) for rice, soy and wheat are provided in the Appendix.}
\label{fig:weights}
\end{figure}

Considering the weight of the links, we can also calculate the average link weight, $\mean{w^{(c)}(y)}=(1/N^{(c)}(y))\sum_{i<j}^{N^{(c)}(y)}w_{ij}^{(c)}(y)=\mean{\mathrm{exp}^{(c)}(y)}$, which is equal to the average total export per country.
Fig. \ref{fig:netwDens}~(b) shows that the average total export in fact decreases for maize and wheat trade and increases for soy trade, while there is no clear trend for rice trade. 
However, the average export is not a suitable measure to describe such trends, because the weight distributions are highly skewed. 
This is shown in Fig. \ref{fig:weights} for two different years, 1992 and 2013, to also allow a comparison of the changes over time.

We note that in all cases the weights are much smaller in 1992, but the distribution is always very broad. 
While the distributions for maize, rice and soy export are right skewed, i.e. have mostly smaller weights, for wheat trade there is a larger fraction of links with big export volumes. 
If we recast the total trade volumes of the different crop in terms of caloric values, we find that the highest amount of calories is \emph{traded} in form of wheat.
Still, the most calories are \emph{produced} in form of maize in 2013.

\section{Modeling the impact of shocks}
\subsection{Dynamics of cascades}
\label{sec:dynamics-cascades}

The main goal of our model is to determine how shocks in the production of one crop in a given country $k$ will affect its availability in other countries $i\in N^{(c)}(y)$.
Such shocks can have different origin as discussed in the Introduction, but we model them here consistently as a  \emph{one time exogenous reduction} $\mathrm{shock}^{c}_{k}$ of the available crop in \emph{one} country. 
Because of the \emph{annual} data, we cannot observe how a country responds to such shocks on a shorter time scale $t$, e.g. within days or weeks, and how such responses affect other countries. 
Here, our model comes into play to proxy such dynamics on the food trade network on the discrete time scale $t=1,2,...,T$, where the maximum time $T$ is less than one year. 

\begin{figure}[htbp]
 \centering
\subfloat{$t$=0}{\includegraphics[width=0.19\textwidth]{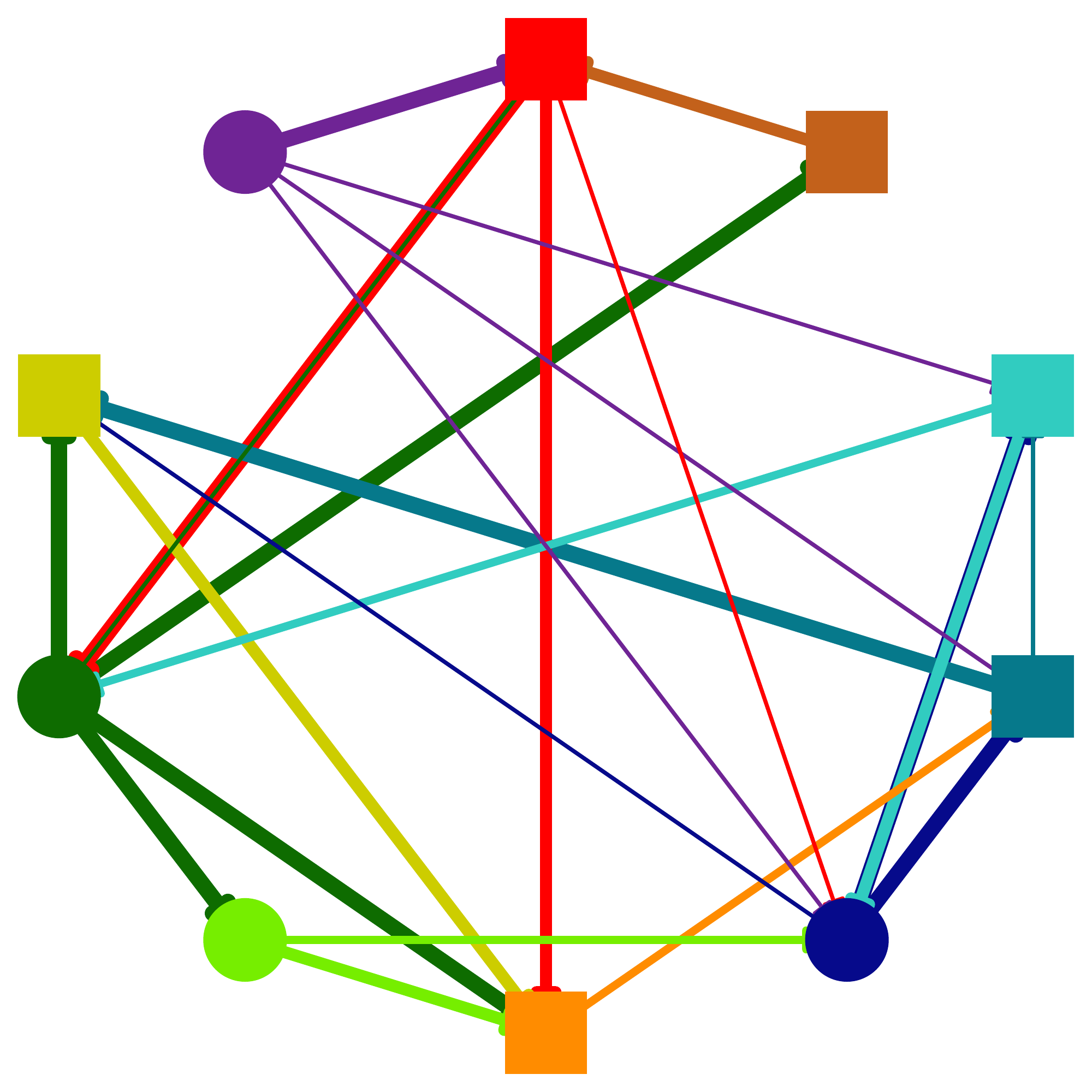}}\hfill 
\subfloat{$t$=1}{\includegraphics[width=0.19\textwidth]{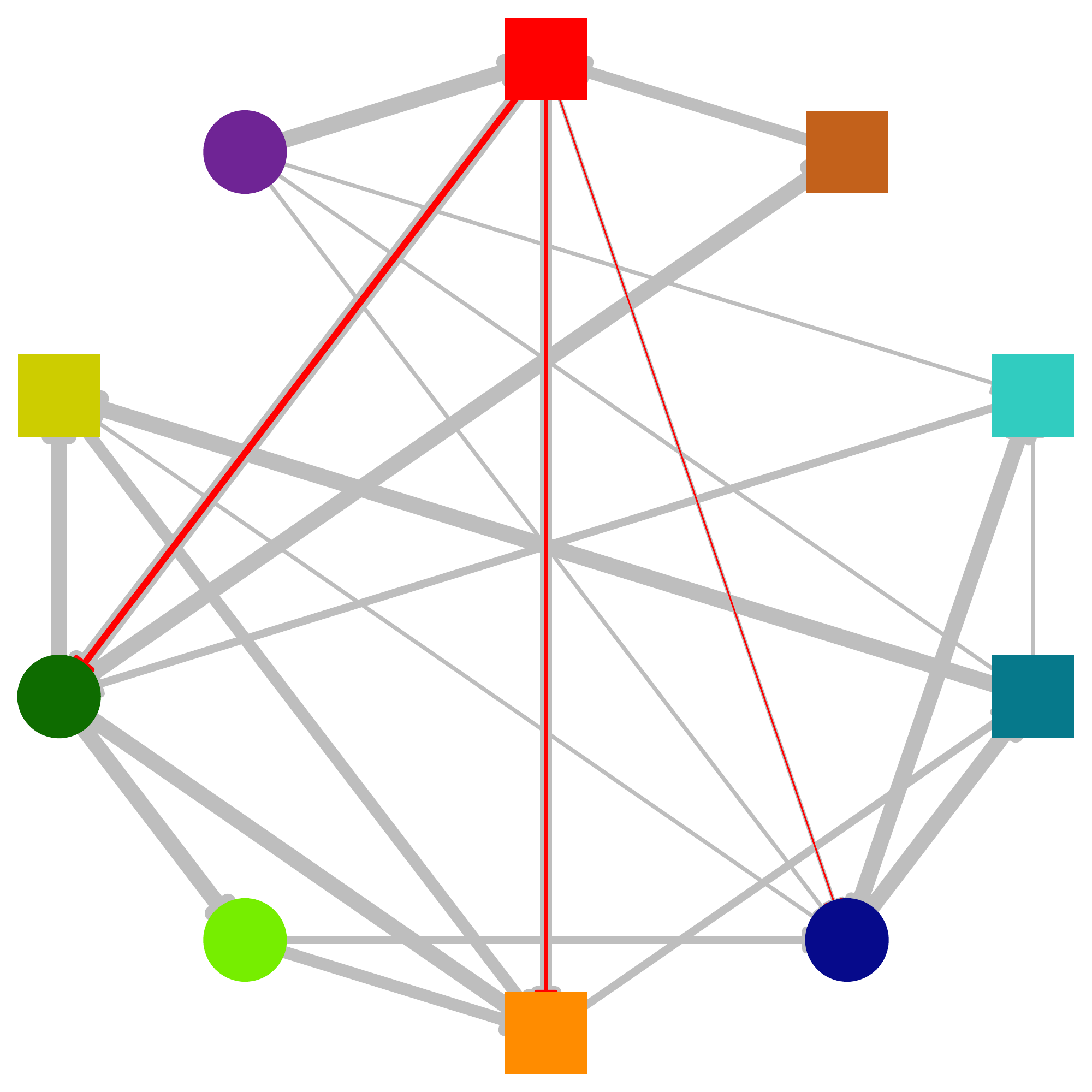}}\hfill \subfloat{$t$=2}{\includegraphics[width=0.19\textwidth]{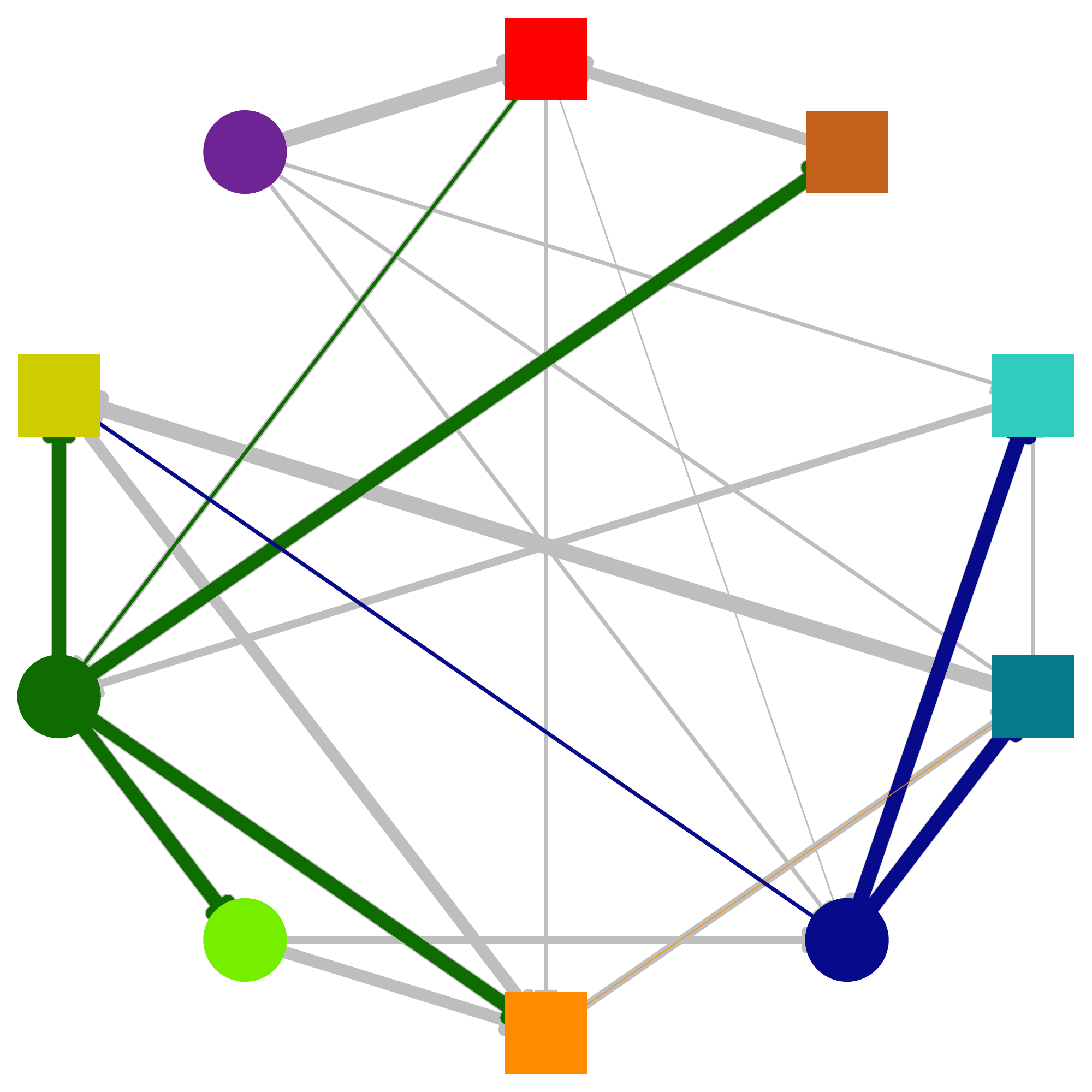}}\hfill \subfloat{$t$=3}{\includegraphics[width=0.19\textwidth]{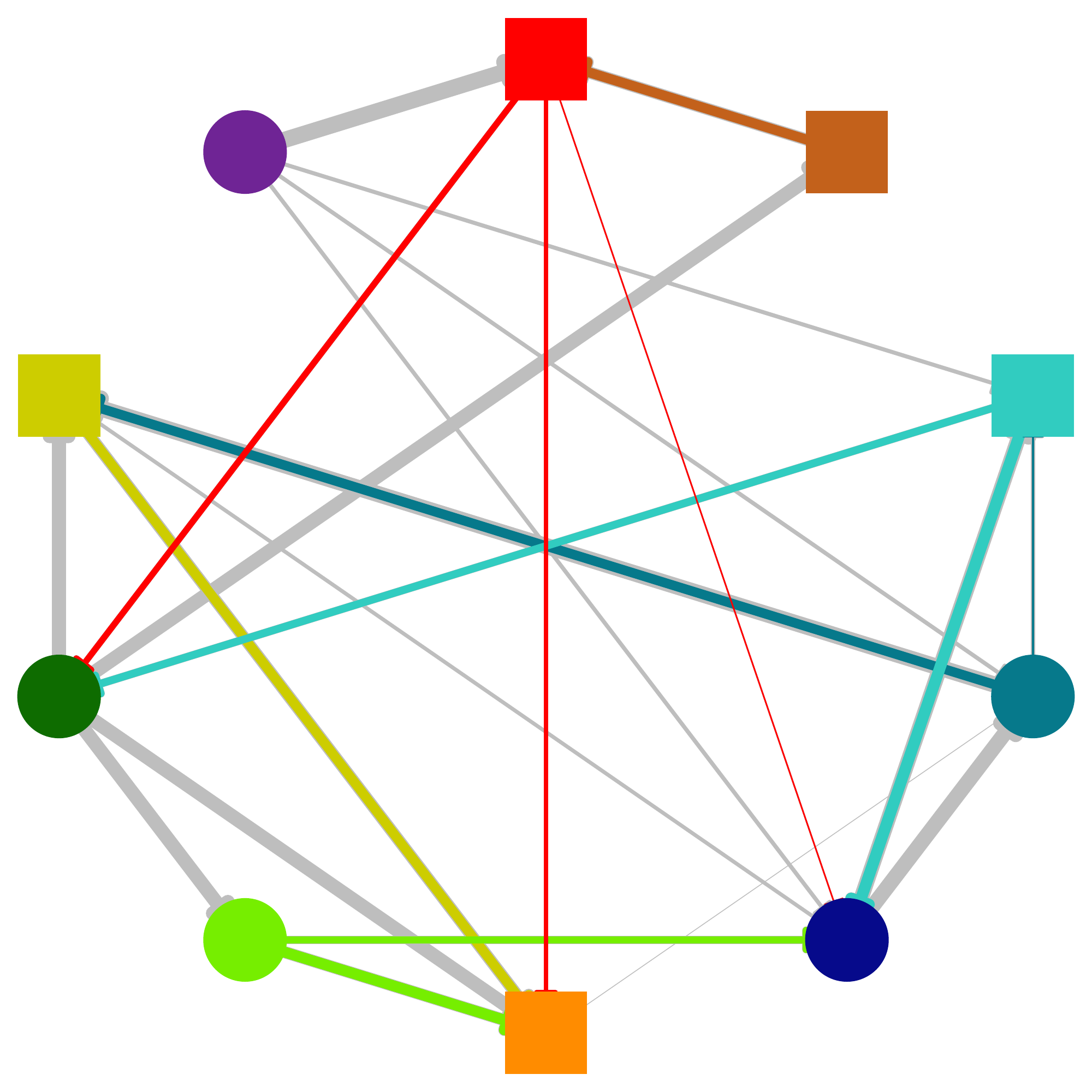}}
\caption{Exemplary cascade process. The previous network is colored in gray, while changing links are colored according to the exporting node. ($t$=0) Initial trade network. ($t$=1) The red node is shocked and reduces its exports. ($t$=2) Importers from the red node compensate for their loss by reducing their own exports. ($t$=3) Further nodes face a demand deficit because of decreased imports and reduce their exports.}
 \label{fig:CascadeModel}
\end{figure}
$t$=0 refers to the reported data at the end of year $y$, i.e. we know for each country $\mathrm{dem}^{(c)}_{i}(t=0)=\mathrm{dem}^{c}_{i}$,  $\mathrm{prod}^{(c)}_{i}(t=0)=\mathrm{prod}^{c}_{i}$, $\mathrm{imp}^{(c)}_{i}(t=0)$, $\mathrm{exp}^{(c)}_{i}(t=0)$.
But theses initial conditions change for every year.
Within one year, we assume that demand and production are fixed to $\mathrm{dem}^{c}_{i}$, $\mathrm{prod}^{c}_{i}$, whereas imports and exports can change on a time scale $t$, i.e. $\mathrm{imp}^{(c)}_{i}(t)$, $\mathrm{exp}^{(c)}_{i}(t)$.
If country $k$ is shocked at $t$=1 by a $\mathrm{shock}^{(c)}_{k}$,  a \emph{demand deficit}
$\mathrm{dd}^{(c)}_{k}(t=1)=\mathrm{shock}^{(c)}_{k}$ will result.
To compensate for that, $k$ reduces its export in the next time step, if possible, such that  $\mathrm{dd}^{(c)}_{k}(t=2)=0$.
This reduction, however, will affect all countries that import the given crop from $k$.
At $t$=2 these countries will face a demand deficit $\mathrm{dd}_{i}(t=2)$ which they try to reduce, this way affecting all other countries that import from them.
Therefore, a cascade resulting from export restrictions evolves in the food trade network on time scale $t$, which involves more and more countries. 
This is illustrated in Fig.~\ref{fig:CascadeModel}.

To formalize the model, we have to express the demand deficit of each country that was not shocked initially: 
\begin{align}
 \mathrm{dd}_{i}(t) = \mathrm{dem}_{i} - \mathrm{prod}_{i} - \mathrm{imp}_{i}(t) + \mathrm{exp}_{i}(t).\label{eq:4}
\end{align}
If $\mathrm{dd}_{i}(t) > 0$, $i$ reduces its exports if possible:
\begin{align}
 \mathrm{exp}_{i}(t+1) = \mathrm{exp}_{i}(t) - \min\left\{\mathrm{dd}_{i}(t), \mathrm{exp}_{i}(t)\right\}.
  \label{eq:6}
\end{align}
Hence, either its deficit vanishes in the next time step, $\mathrm{dd}_{i}(t+1) = 0$, or at least all current exports are stopped.

To complete Eq.~\eqref{eq:4}, we have to specify how $\mathrm{imp}_{i}(t)$ is affected by the export reductions of other countries.
According to Eq.~\eqref{eq:1}, imports are defined through the weights $w_{ji}^{(c)}(t)$ which will change on time scale $t$ if export restrictions occurred. 
We assume that exporting countries do not change their preference for specific countries at the short time scale $t$.
I.e., in case of an export reduction every of their importers is proportionally affected. 
This implies that the ratio $w_{ji}(t)/\mathrm{exp}_{j}(t) = w_{ji}(0)/\mathrm{exp}_{j}(0)$ is constant over $t$ and can be set to the initial value,  where the initial trades $w_{ji}(0)$ are entries of the trade matrix $W$ for a given year.
This gives for  the dynamics of the trade weights
\begin{align}
  w_{ji}(t) =  \mathrm{exp}_{j}(t) \frac{w_{ji}(0)}{\mathrm{exp}_{j}(0)}\;;\quad \mathrm{imp}_{i}^{(c)}(t) = \sum^N_{j=1} w_{ji}^{(c)}(t)
  \label{eq:7}
\end{align}
Given an initial shock $\mathrm{shock}_{k}$, the combined equations \eqref{eq:4},\eqref{eq:6},\eqref{eq:7} determine the dynamics of the cascade.
The final step of the cascade at time $t=T$ is reached if no country with a demand deficit can further reduce its export.
This usually applies to more than one country because the cascade has evolved along various paths, determined by the number of importers.
How many countries are eventually left with a non-reducible demand deficit thus depends on the initial country that could be an important producer, the size of the shock, but also on the sequence in which countries are involved.  
Hence, in order to systematically study such effects, we need an approach that does not just consider a single event.
This is developed in the following.

\subsection{Shock scenarios}

To assess the vulnerability of the trade network, we consider two different types of shocks that each represent a different limit case:
An \emph{equal shock} generates a \emph{fixed} demand deficit of the shocked country, no matter whether this is a small or a large country.
This allows us to study how the \emph{same} deficit would affect different countries. 
To define the size of the schock, we set this to  25\% of the production of an average country, $\mathrm{shock}^{(c)}_{k}=0.25 \mean{\mathrm{prod}^{(c)}(y)}$ in a given year $y$.
Only if the size of the shock exceeds the shocked country's production $\mathrm{prod}^{(c)}_{k}(y)$ and demand $\mathrm{dem}^{(c)}_k(y)$, we limit $\mathrm{shock}^{(c)}_k(y)$ to the maximum of both:
$\mathrm{shock}^{(c)}_{k} = \max\left(\mathrm{prod}^{(c)}_k(y), \mathrm{dem}^{(c)}_k(y)\right)$.
    
The second type of shocks, at difference with the first one, is  not equal for all countries but proportional to the production or demand of the shocked country, i.e. $\mathrm{shock}^{(c)}_{k} = 0.25\,\max\left(\mathrm{prod}^{(c)}_k(y), \mathrm{dem}^{(c)}_k(y)\right)$. 
Therefore, we call this a \emph{proportional shock}.
It allows us to study how countries with very different production impact the size of the cascades.
Proportional shocks of 25\% can be seen as quite large.
However, our data shows that they have happened in more than 5\% of all changes of production and demand for all countries and years.
I.e., proportional shocks of this size are \emph{not} negligible, but realistic.

If we apply a shock to a given country $k$, we will observe cascades of export restrictions as illustrated in Fig.~\ref{fig:CascadeModel}.
The outcome characterizes the influence only of country $k$, thus, we have to run the model with every possible country $k=1,...,N^{(c)}(y)$ as the target of a shock.
In order to visualize the influence of all countries together, we generate  a \emph{higher-order trade dependency network} as explained in the following.

\section{Results}\label{sec:results}

\begin{figure}[htbp]
  \centering
   \subfloat{(a)}{\includegraphics[width=0.45\textwidth]{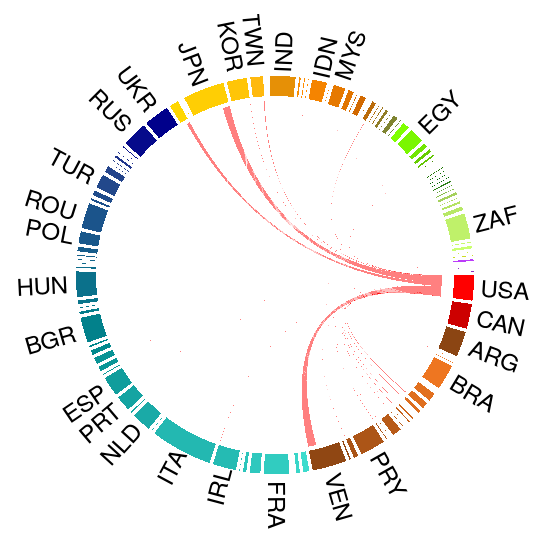}}
  \hfill
   \subfloat{(b)}{\includegraphics[width=0.45\textwidth]{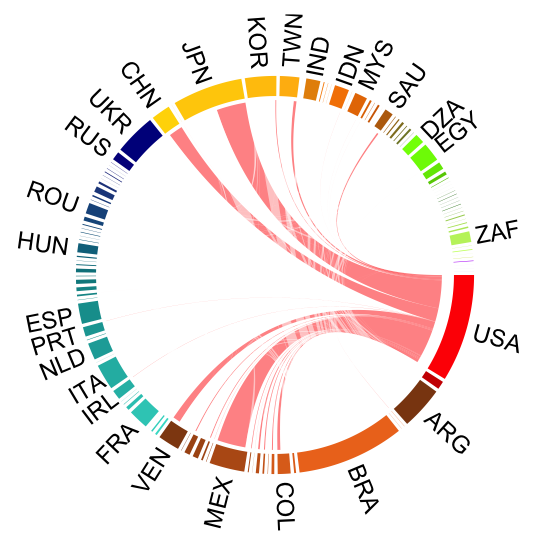}}
  \caption{An initial shock of the production of maize in the USA in 2013 eventually causes a cascade (not shown) that lead to demand deficit in  countries that are no direct trade partners of the US. (a) Equal shock, (b) proportional shock scenario. The link strength is proportional to the demand deficit caused. Note that only links with large weights are visible.}
  \label{fig:usa}
\end{figure}

Our aim is to visualize the final outcome for the collection of cascade processes starting in \emph{all} countries. 
Fig. \ref{fig:usa} explains this procedure for the one-time exogenous shock of a \emph{single} country, the USA, in 2013, only for maize trade. 
All links start in the shocked country, the US, and end in different countries which all face a demand deficit at the end of a cascade. 
The link strength   is proportional to this deficit. 
We do not show the intermediate steps, only the final outcome, i.e. each link connects the origin of a cascade with a number of finally affected countries.

Fig. \ref{fig:usa} allows to compare the influence of an equal shock (a) with that of a proportional shock (b) of the main producer of maize.
Because of the large production, the \emph{proportional} shock of the US is larger than for the \emph{equal} shock.
Therefore, it inflicts higher demand deficits also in more countries.
The difference, however, does not scale with the shock size, due to the nonlinear cascades.
For instance, Mexico (MEX) is significantly affected by the proportional shock, but not at all by the equal shock.
Japan (JPN), on the other hand, faces a high demand deficit even for smaller shocks of the US.

In order to visualize the influence of all countries together, we generate  a \emph{higher-order trade dependency network} by 
combining  the final outcomes of cascades for all possible countries $k=1,...,N^{(c)}(y)$ as starting points.
A zero-order network would simply be the empirically observed trade network shown in Fig. \ref{fig:data}. 
The first-order network would show the impact of shocks on direct export partners, the second-order network the impact on the export partners of those direct partners, i.e. the indirect impact with respect to the initially shocked country, etc. 
The highest order is given by the maximum number of steps in all cascades. 
Hence, higher-order dependency networks reflect the ability of countries to compensate demand deficits by export restrictions.

\begin{figure}[htbp]
  \centering
 \subfloat{(a)}{\includegraphics[width=0.45\textwidth]{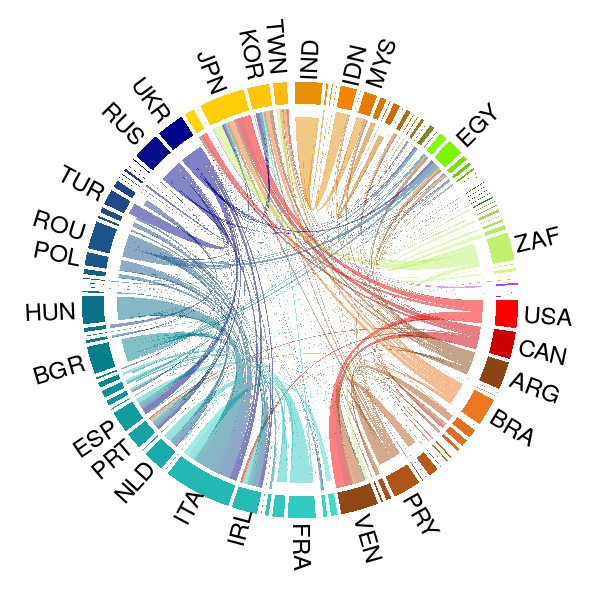}}\hfill\subfloat{(b)}{\includegraphics[width=0.45\textwidth]{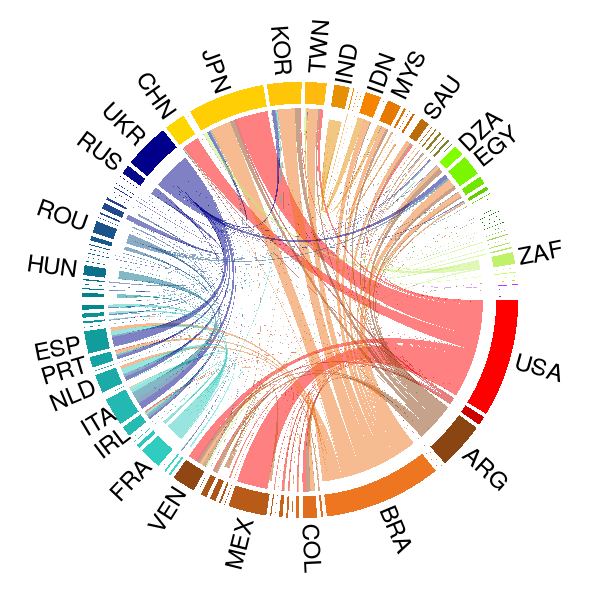}}
 \caption{The higher-order trade dependency network for maize in 2013.
  The bar size of each country represents its weighted degree, i.e. the sum of all in-coming and out-going link weights.
  (a) Equal shock, (b) proportional shock scenario.
For the visualizations, we use the circlize package in R \cite{krzywinski-circos-2009}. 
}
  \label{fig:shockNetwComp}
 \end{figure}

The higher-order trade dependency network for maize is shown in  Fig. \ref{fig:shockNetwComp}~(a,b) for equal shocks and for proportional shocks.
We note that proportional shocks emphasize the impact of the biggest producers, in particular USA, Brazil, and Argentina. 
Interestingly, shocks in the USA finally impact many countries in South America, while shocks in Brazil mostly impact Asia, but also Africa. 
Equal shocks, on the other hand, highlight dependencies in general, not just on the biggest producers.
Shocks of European countries mostly impact other European countries, with Italy as the most affected country.
With respect to Africa, a shock of South Africa generates the highest demand deficit not in Africa, but in Japan and in South America.

\begin{figure}[htbp]
 \centering
\subfloat{(a)}{\includegraphics[width=0.3\textwidth]{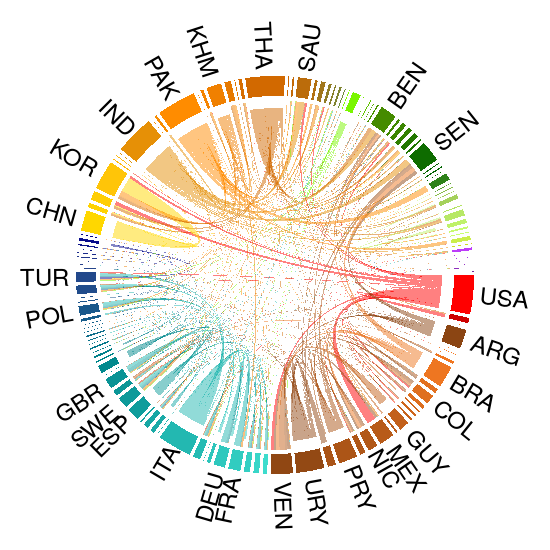}}\hfill
 \subfloat{(b)}{\includegraphics[width=0.3\textwidth]{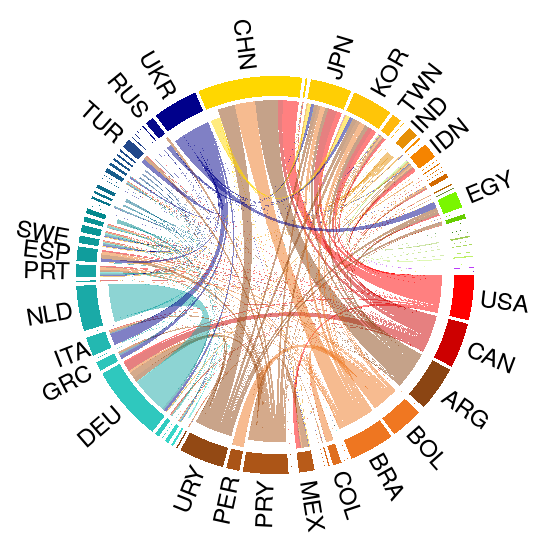}}\hfill
 \subfloat{(c)}{\includegraphics[width=0.3\textwidth]{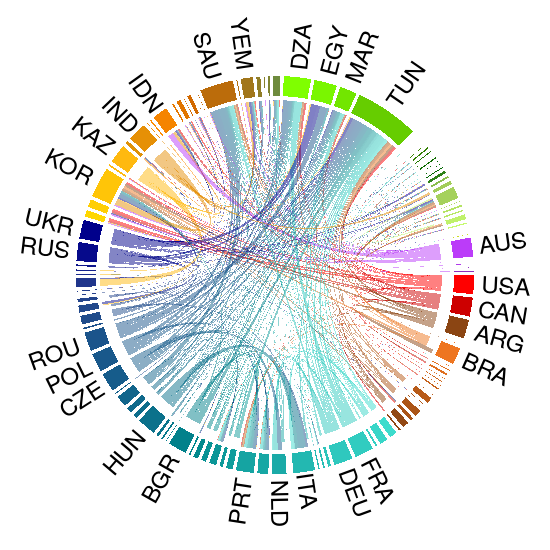}}
\subfloat{(d)}{\includegraphics[width=0.3\textwidth]{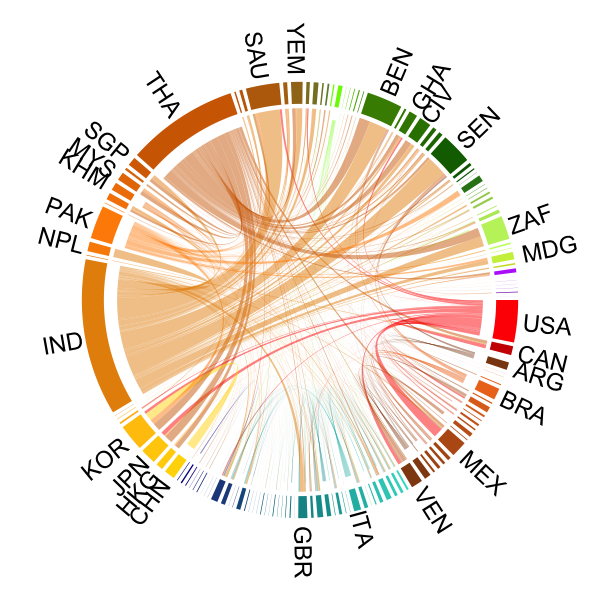}}\hfill
 \subfloat{(e)}{\includegraphics[width=0.3\textwidth]{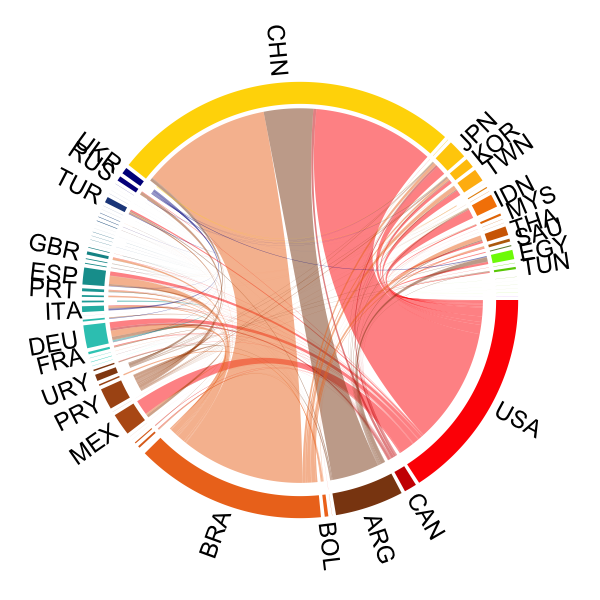}}\hfill
 \subfloat{(f)}{\includegraphics[width=0.3\textwidth]{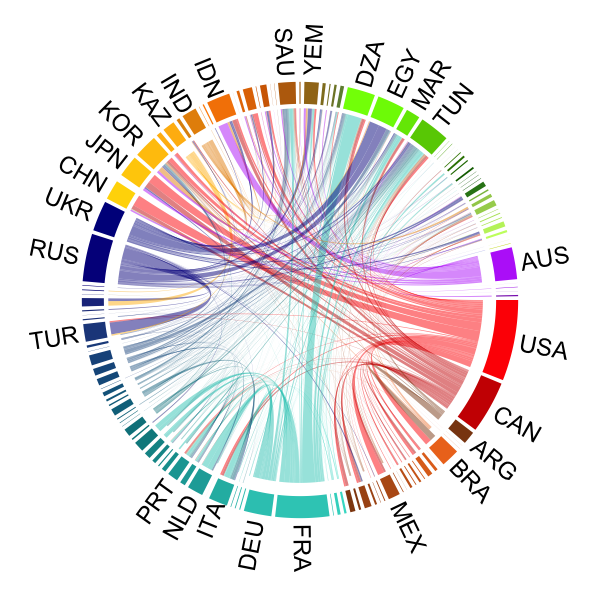}}

 \caption{Cascade dependencies in 2013 for (a/d) rice, (b/e) soy, and (c/f) wheat trade. (top row)  Equal shock, (bottom row) proportional shock scenario.
}
 \label{fig:OLRDnetwork}
\end{figure}

The higher-order trade dependency network of maize can be also compared with the respective networks for rice, soy and wheat shown in Fig.~\ref{fig:OLRDnetwork}(a,b,c) both for equal (top) and proportional shock (bottom) scenarios.
It is apparent that shocks of the US, for all crops and all scenarios, have a major impact on the demand deficit of other countries, even for \emph{rice}.
Equal shock scenarios highlight the importance of European countries as intermediaries.
I.e., they import, add value, and export. 
In particular, shocks of wheat production and demand in European countries affect the whole world as shown by the rather homogeneous link distribution.  

Looking at the impact of the biggest producers of \emph{soy}, we find that the strongest dependencies are between the US, Brazil and Argentina on the one hand, and China on the other hand.
The higher-order trade dependency network appears to be similar to the one of maize (Fig.~\ref{fig:shockNetwComp}b) as both maize and soy share the main producers.
On the other hand, both the zero-order (Fig. \ref{fig:data}) and the higher-order trade dependency network of soy are less dense than the ones for maize. 
In consequence, the impact of shocks is more concentrated on a few countries.

Regarding rice, we observe the prominent role of Asian countries.
Shocks of India or Thailand appear to have the most critical impact on other countries, in particular in Africa.
Surprisingly, also shocks of the USA are relevant for a few Asian countries like Japan and Korea.

Africa mainly depends on rice and wheat imports from Asian and European countries, as illustrated by the fact that the incoming links in the higher-order trade dependency network have a color different from the African countries. 

\section{Discussion}
\label{sec:discussion}

In this paper, we provide a quantitative analysis of the global food trade of the four major internationally traded  crops, maize, rice, soy, and wheat.
This analysis makes two major contributions.
First, from the available data we have reconstructed the global trade network between 176 countries for 21 years and have evaluated network properties such as link density and distribution of link weights over time.
This is complemented by an empirical analysis of the production, import, export and resulting demand of each country for each year.
We show that the roles of countries in the global food trade cannot be separated, i.e., many counties are producers, importers and exporters of the same crop at the same time.
This highlights the importance of countries as intermediaries, either for trade or food processing, and points to the increasing complexity of global value chains.

This insight has motivated our second major contribution, a model to reveal the indirect dependencies between countries if the the production or demand for food in a specific country was shocked exogeneously (e.g. by natural desasters).
The model reflects that countries can compensate for shortages in the supply of a given staple food by imposing \emph{export restrictions}.
These impact their direct trade partners, which try to compensate supply deficits as well by export restrictions.
This way, cascades emerge on the given trade network which involve many countries.
They only stop if a country cannot further compensate its demand deficit by export restrictions.
Our cascade model captures a process that cannot be observed given the available data.
That means, it allows to relate the country shocked originally with the country that eventually suffers the most from this shock, at the end of a cascade.
These indirect relations are neither obvious, nor have they been revealed in the existing literature.

To reflect these dependencies on the global level, we have developed \emph{higher-order trade dependency networks}.
They visualize the impact of an initial shock in \emph{any} country on those countries that eventually get a demand deficit from this shock.
We have calculated these visualizations for two different shock scenarios.
Proportional shocks highlight the impact of big producers, while equal shocks highlight general dependencies.  
The value of these higher-order trade dependency networks is in revealing indirect dependencies between countries.
As food trade becomes more complex and more countries become involved (see Fig. \ref{fig:N}), disentangling a country's impact on the globalized food trade gives valuable information also for policy makers.
They are enabled to anticipate how shocks of different sizes in a given country impact other countries directly \emph{and} indirectly, via export restrictions. 

In summary, we could quantify that a great number of Asian and African countries are most exposed to cascades.
Noticeably, the main suppliers are similar for most of the crops: USA, Canada, Argentina, Brazil, and India.
While shocks in the USA mainly affect South America and several Asian countries, the south of Africa is primarily dependent on American and Asian exporters. 
The north of Africa depends strongly on Europe, in particular via wheat imports.
Remarkably, a great number of European countries appear frequently among the main trade intermediaries.

While our results obtained already allow for policy considerations, we mention two aspects that should be kept in mind.
The first one regards substitutions of different staple food.
In our model, demand deficits are considered as independent across the considered crops maize, rice, soy, wheat, whereas in practical situations a shortage of one food may be compensated, at least partially, by other food.
Such couplings between cascades could be considered, but need to be based on justified assumptions.
These have to also include prices which are currently left out in our model.
This is the second aspect to be kept in mind.

In principle, price dynamics could affect the dynamics and the outcome of cascades because both supply and demand depend on price \cite{pricePNAS,exportban}.
Therefore, studies on seafood trade \cite{seafood} have considered decreasing demand due to price increases.
We argue that the situation is different for staple food because its \emph{price elasticity} is small, i.e. the demand does not change substantially despite increasing prices.
This reflects that staple food is a basic need for most of the population to exist.
Substitution of one crop by other (cheaper) crops may be an alternative only for less processed products. 
However, this leads to joined price movements of the alternatives. so that the overall demand for one crop should not change drastically. 
Eventually, for highly processed products the price of raw ingredients constitute only a smaller share 
thus price changes do not alter the consumer price significantly.

Our model emphasizes the impact of export restrictions rather than price fluctuations in response to an external shock.  
According to our analysis, especially rice trade is prone to cascading export restrictions. 
Unfortunately, those cascades are empirically most frequently observed \cite{exportban}, 
but also other crop trade is vulnerable to cascading export restrictions. 
At the same time, cascades can provide means for good shock diversification. 
To which extend food losses are critical for specific countries is decided by the amount of national stocks available for compensation.
Our modeling approach provides a way to proxy the necessary size of such stocks.

\subsection*{Acknowledgements}
\label{sec:Acknowledgements}

RB acknowledges support by the ETH48 project of the ETH Risk Center. RB thanks the participants of the conference 'Tackling World Food System Challenges: Across Disciplines, Sectors, and Scales` in 2015 for fruitful discussions.

\normalsize
\begin{appendix}

\section*{Appendices}
  
\section{Available data and inconsistencies}
\label{sec:network-construction}

The detailed trade matrix dataset provided by the Food and Agricultural Organization of the United Nations \cite{dataTrade} distinguishes between reported exports and imports.

In some cases, the reports of a trade as export and as import are not consistent. 
We take a conservative approach by regarding only the minimum of both trades.
This way, we underestimate the total international trade. 
According to the rule of thumb that a high network connectivity and thus high international trade support long cascades, we usually tend to underestimate the severity of cascades, but also their possible shock diversification effect.
An alternative approach would be to take the average between the reported export and import, as for instance implemented by \cite{Puma2015}. 
In this case, we could not guaranty that we over- or underestimate the international trade connectivity.
However, both network construction approaches lead to qualitatively similar results.

In consequence of the minimal trade approach, we only consider reporting countries by the FAO. 
A full list of all countries that engage at least once in the trade or production of a crop between 1992 and 2013 and their ISO-3 code is given by Table \ref{table:ctryList}.
Although the international borders have been rather stable from 1992-2013, we still have to handle a few changes. Since Belgium and Luxembourg form an economic union and their trade statistics are only available for the combination of both 
till 1999, we merge their data in our whole analysis to provide consistency and acknowledge their economic union. 
Czechoslovakia was divided into Czech Republic and Slovakia in 1993, so rather in the beginning of our observation period.
So, we keep both countries separated in our analysis (and simply assign the 1992 value to one of the countries without any consequence). 
Yugoslavia (1992-2003) and the state union of Serbia and Montenegro (2003-2006) was split into Montenegro and Serbia. 
We regard them as one entity till 2005, and afterwards as separate. 

\scriptsize
\begin{longtable}{| p{.08\textwidth} | p{.36\textwidth} | p{.08\textwidth} | p{.36\textwidth} |} 
\hline
ISO 3166-1 alpha-3 codes & Official country name & ISO 3166-1 alpha-3 codes & Official country name\\ 
  \hline
USA & United States of America & TWN & Taiwan, the Republic of China \\ 
  CAN & Canada & KAZ & Republic of Kazakhstan \\ 
  ATG & Antigua and Barbuda & KGZ & Kyrgyz Republic \\ 
  ARG & Argentine Republic & AFG & Islamic State of Afghanistan \\ 
  BHS & Commonwealth of the Bahamas & BGD & People's Republic of Bangladesh \\ 
  BRB & Barbados & BTN & Kingdom of Bhutan \\ 
  BMU & Bermuda & LKA & Democratic Socialist Republic of Sri Lanka \\ 
  BOL & Republic of Bolivia & IND & Republic of India \\ 
  BRA & Federative Republic of Brazil & IRN & Islamic Republic of Iran \\ 
  ABW & Aruba & MDV & Republic of Maldives \\ 
  BLZ & Belize & NPL & Kingdom of Nepal \\ 
  CHL & Republic of Chile & PAK & Islamic Republic of Pakistan \\ 
  COL & Republic of Colombia & BRN & Negara Brunei Darussalam \\ 
  CRI & Republic of Costa Rica & IDN & Republic of Indonesia \\ 
  CUB & Republic of Cuba & KHM & Kingdom of Cambodia \\ 
  DMA & Commonwealth of Dominica & MYS & Malaysia \\ 
  ECU & Republic of Ecuador & PHL & Republic of the Philippines \\ 
  SLV & Republic of El Salvador & SGP & Republic of Singapore \\ 
  GRD & Grenada & THA & Kingdom of Thailand \\ 
  GTM & Republic of Guatemala & BHR & State of Bahrain \\ 
  GUY & Cooperative Republic of Guyana & QAT & State of Qatar \\ 
  HND & Republic of Honduras & SAU & Kingdom of Saudi Arabia \\ 
  JAM & Jamaica & OMN & Sultanate of Oman \\ 
  MEX & United Mexican States & ARE & the United Arab Emirates \\ 
  MSR & Montserrat & YEM & Republic of Yemen \\ 
  NIC & Republic of Nicaragua & JOR & Hashemite Kingdom of Jordan \\ 
  PAN & Republic of Panama & KWT & State of Kuwait \\ 
  PRY & Republic of Paraguay & LBN & Lebanese Republic \\ 
  PER & Republic of Peru & SYR & Syrian Arab Republic \\ 
  KNA & Federation of Saint Kitts and Nevis & ISR & State of Israel \\ 
  LCA & Saint Lucia & DZA & People's Democratic Republic of Algeria \\ 
  VCT & Saint Vincent and the Grenadines & EGY & Arab Republic of Egypt \\ 
  SUR & Republic of Suriname & LBY & Great Socialist People's Libyan Arab Jamahiriya \\ 
  TTO & Republic of Trinidad and Tobago & MAR & Kingdom of Morocco \\ 
  URY & Oriental Republic of Uruguay & SDN & Republic of the Sudan \\ 
  VEN & Bolivarian Republic of Venezuela & TUN & Republic of Tunisia \\ 
  AUT & Republic of Austria & CMR & Republic of Cameroon \\ 
  DNK & Kingdom of Denmark & CPV & Republic of Cape Verde \\ 
  FRO & Fxoroyar (Faroe Is.) & CAF & Central African Republic \\ 
  FIN & Republic of Finland & COG & Republic of the Congo \\ 
  FRA & French Republic & GAB & Gabonese Republic \\ 
  DEU & Federal Republic of Germany & STP & Democratic Republic of Sao Tome and Principe \\ 
  GRC & Hellenic Republic & COD & Democratic Republic of the Congo \\ 
  ISL & Republic of Iceland & BEN & Republic of Benin \\ 
  IRL & Ireland & GMB & Republic of the Gambia \\ 
  ITA & Italian Republic & GHA & Republic of Ghana \\ 
  MLT & Republic of Malta & GIN & Republic of Guinea \\ 
  NLD & Kingdom of the Netherlands & CIV & Republic of Cote D'Ivoire \\ 
  NOR & Kingdom of Norway & MLI & Republic of Mali \\ 
  PRT & Portuguese Republic & MRT & Islamic Republic of Mauritania \\ 
  ESP & Kingdom of Spain & NER & Republic of Niger \\ 
  SWE & Kingdom of Sweden & NGA & Federal Republic of Nigeria \\ 
  CHE & Swiss Confederation & SEN & Republic of Senegal \\ 
  GBR & United Kingdom of Great Britain and Northern Ireland & SLE & Republic of Sierra Leone \\ 
  GRL & Greenland & TGO & Togolese Republic \\ 
  BELLUX & the Kingdom of Belgium and the Grand Duchy of Luxembourg combined & BFA & Burkina Faso \\ 
  REU & Réunion & BDI & Republic of Burundi \\ 
  ALB & Republic of Albania & KEN & Republic of Kenya \\ 
  BGR & the Republic of Bulgaria & RWA & Rwandese Republic \\ 
  CYP & Republic of Cyprus & UGA & the Republic of Uganda \\ 
  EST & Republic of Estonia & ETH & Federal Democratic Republic of Ethiopia \\ 
  BIH & Bosnia and Herzegovina & BWA & Republic of Botswana \\ 
  HUN & Republic of Hungary & MWI & Republic of Malawi \\ 
  HRV & Republic of Croatia & NAM & Republic of Namibia \\ 
  LVA & Republic of Latvia & ZWE & Republic of Zimbabwe \\ 
  LTU & Republic of Lithuania & ZAF & Republic of South Africa \\ 
  MKD & Republic of Macedonia & SWZ & Kingdom of Swaziland \\ 
  CZE & Czech Republic & TZA & United Republic of Tanzania \\ 
  POL & Republic of Poland & ZMB & Republic of Zambia \\ 
  ROU & Romania & COM & Federal Islamic Republic of the Comoros \\ 
  SVN & Republic of Slovenia & MDG & Republic of Madagascar \\ 
  SVK & Slovak Republic & MUS & Republic of Mauritius \\ 
  TUR & Republic of Turkey & SYC & Republic of Seychelles \\
  SRB & Republic of Serbia & AUS & Commonwealth of Australia \\ 
  MNE & Montenegro & SLB & Solomon Islands \\ 
  ARM & Republic of Armenia & COK & the Cook Islands \\ 
  AZE & Republic of Azerbaijan & FJI & Republic of the Fiji Islands \\ 
  BLR & Republic of Belarus & PYF & Territory of French Polynesia \\ 
  GEO & Georgia & KIR & Republic of Kiribati \\ 
  MDA & Republic of Moldova & NCL & Territory of New Caledonia and Dependencies \\ 
  RUS & Russian Federation & VUT & Republic of Vanuatu \\ 
  UKR & Ukraine & NZL & New Zealand \\ 
  CHN & People's Republic of China & PNG & Independent State of Papua New Guinea \\ 
  HKG & Hong Kong Special Administrative Region & TON & Kingdom of Tonga \\ 
  JPN & Japan & TUV & Tuvalu \\ 
  KOR & Republic of Korea & GUF & French Guiana \\ 
  MAC & Macau Special Administrative Region & GLP & Guadeloupe \\ 
  MNG & Mongolia & MTQ & Martinique \\ 
   \hline
 \caption[List of countries considered in our analysis.]{List of countries considered in our analysis.}
  \label{table:ctryList}
\end{longtable}
\normalsize

\clearpage

\section{Production, import, export and demand of rice, soy and wheat across countries}
\label{sec:prod-import-export}

\begin{figure}[htbp]
 \centering
\includegraphics[width=0.49\textwidth]{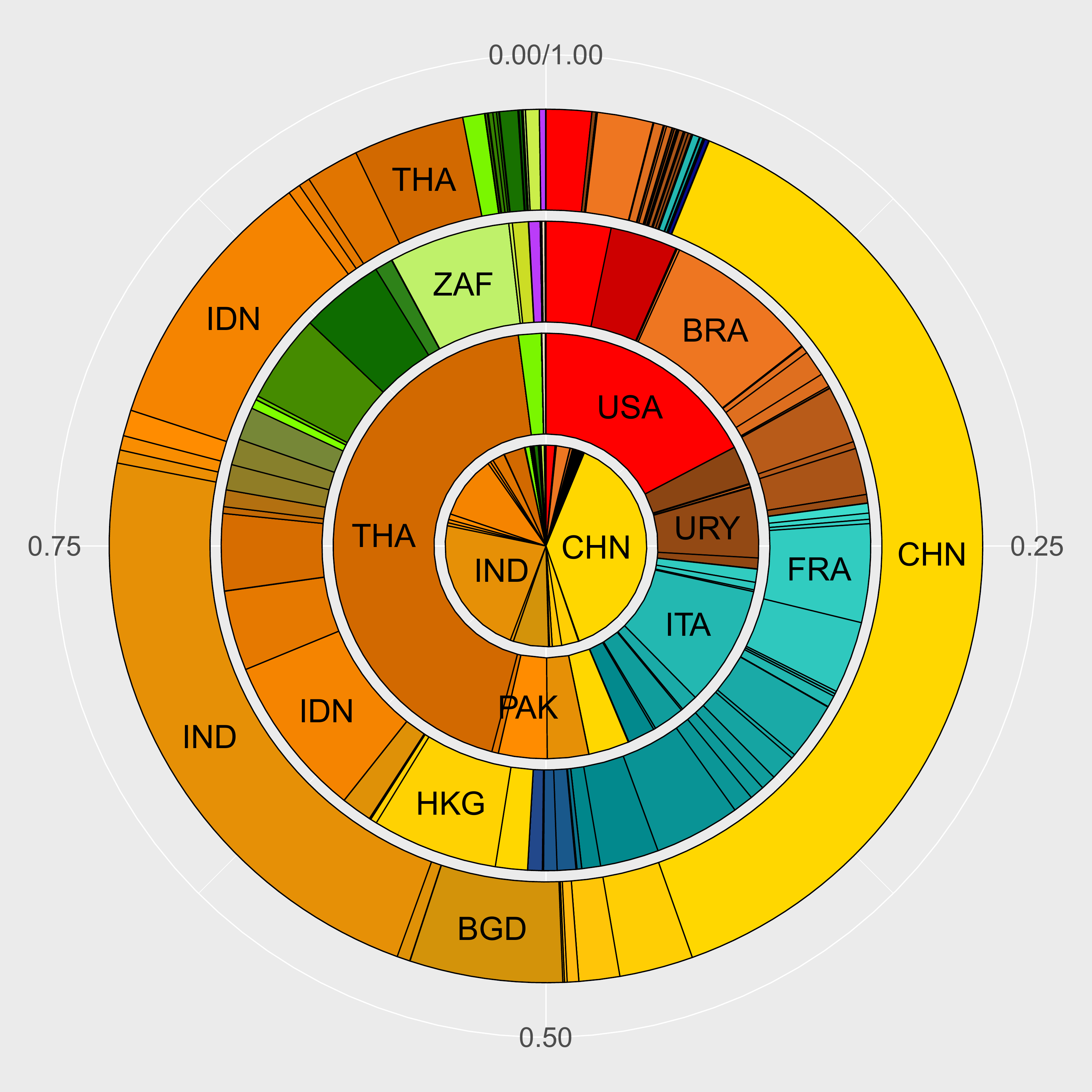}
\includegraphics[width=0.49\textwidth]{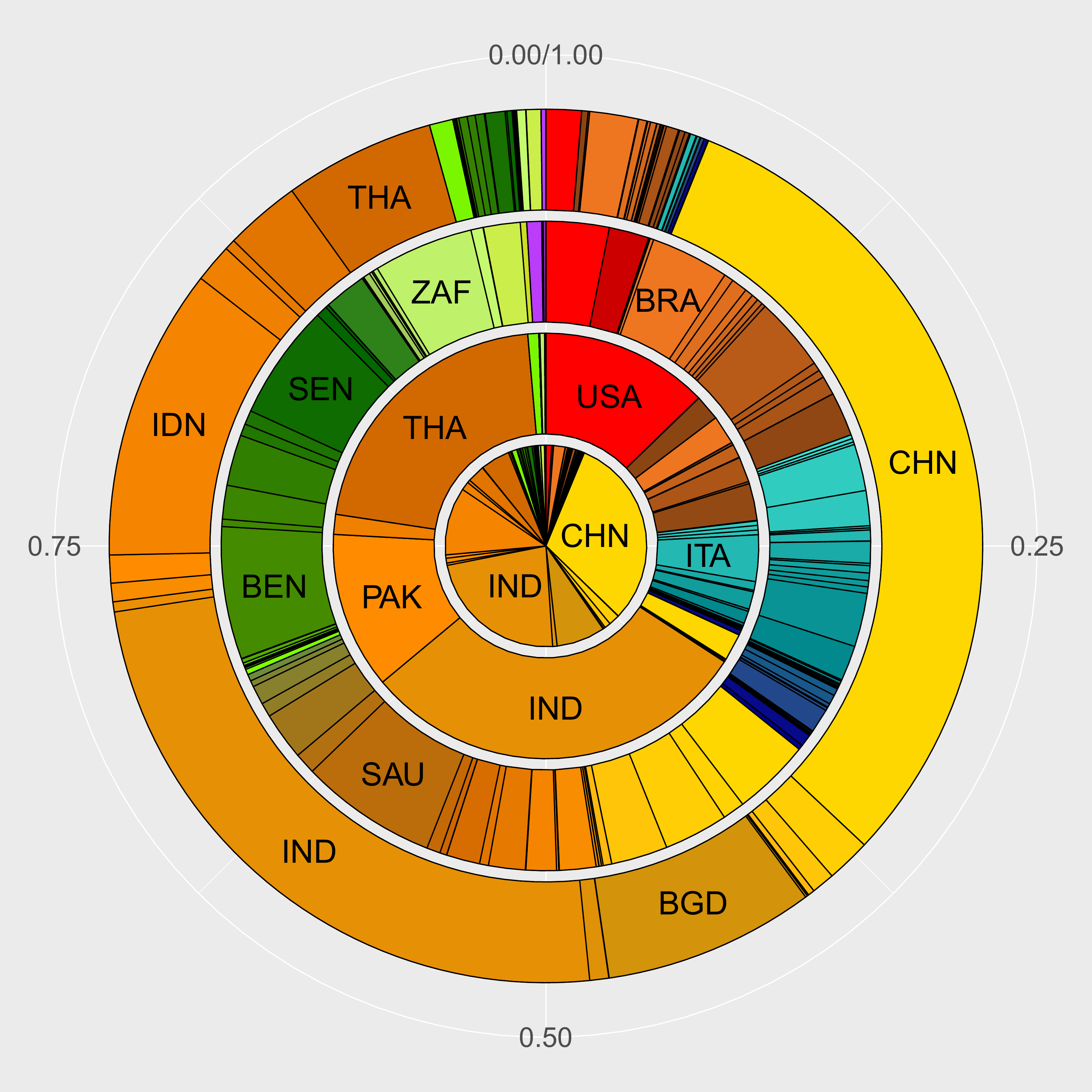}
   \caption{\label{fig:pier} Fractions of rice \emph{production} $\mathrm{prod}_{i}^{(R)}(y)$ (outer circle), \emph{import} $\mathrm{imp}_{i}^{(R)}(y)$ (second outer circle), \emph{export} $\mathrm{exp}_{i}^{(R)}(y)$ (second inner circle) and \emph{demand} $\mathrm{dem}_{i}^{(R)}(y)$ (inner circle) per country in $y$=1992 (left) and $y$=2013 (right).
   }
 \end{figure}

\begin{figure}[htbp]
 \centering
\includegraphics[width=0.49\textwidth]{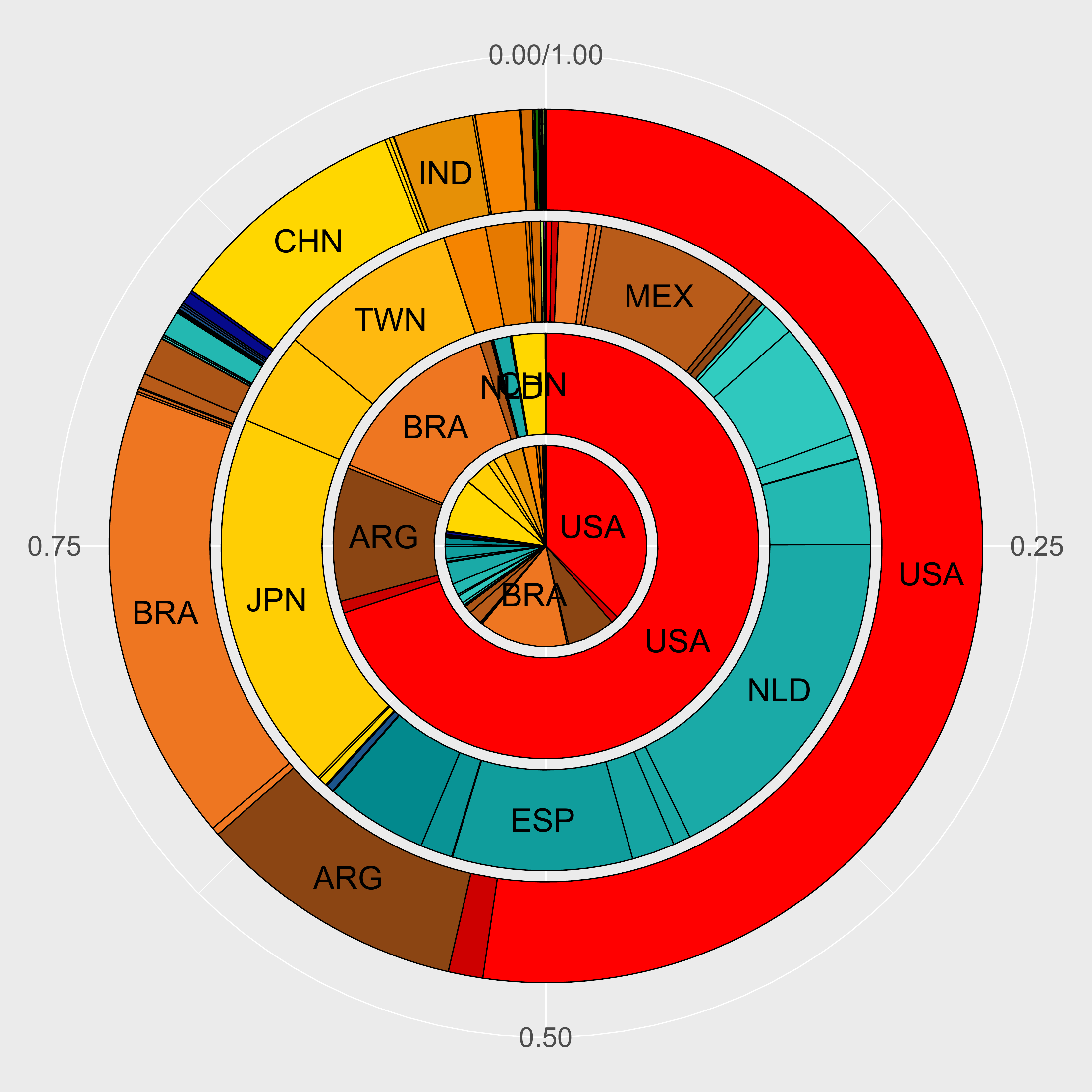}
\includegraphics[width=0.49\textwidth]{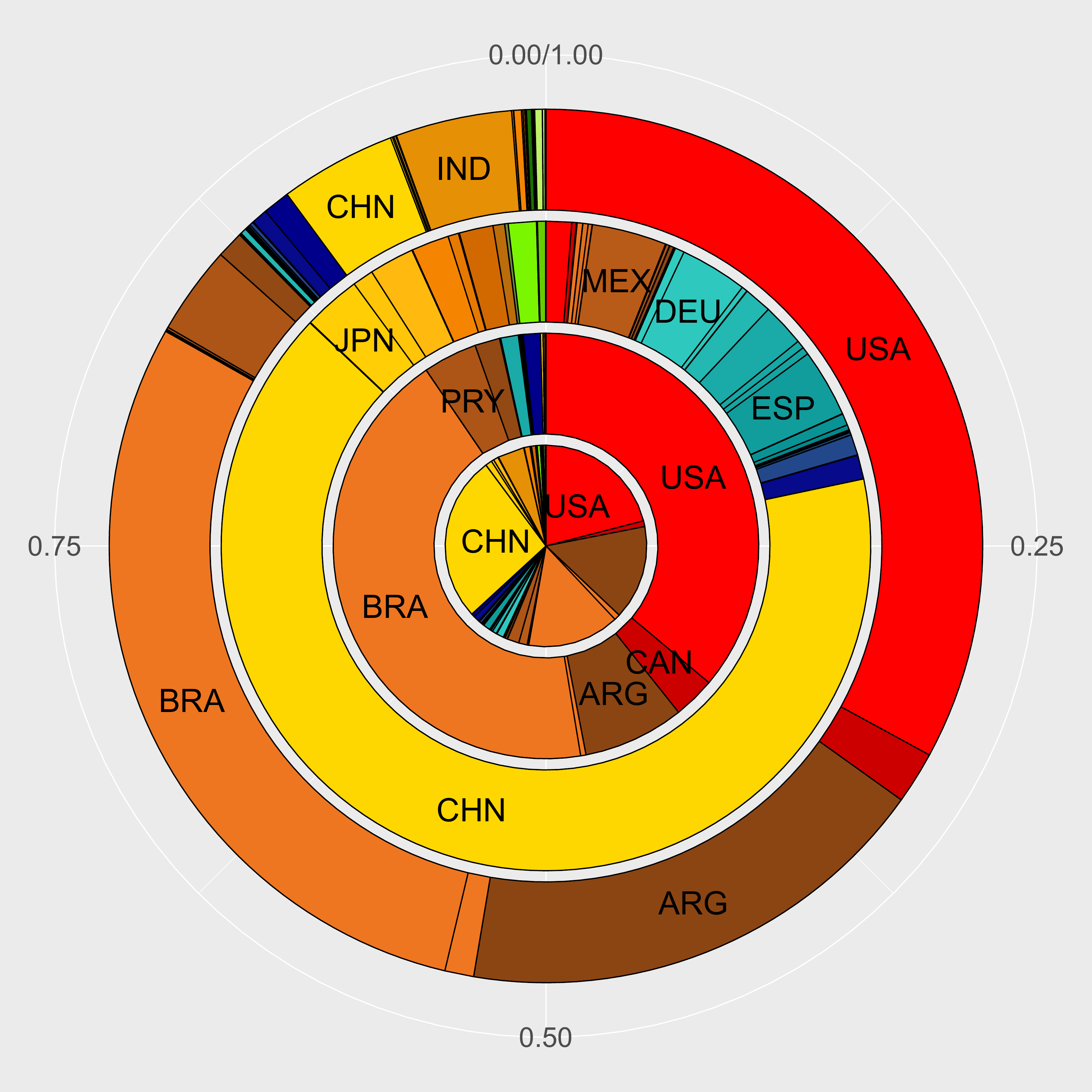}
   \caption{\label{fig:pies} Fractions of soy \emph{production} $\mathrm{prod}_{i}^{(S)}(y)$ (outer circle), \emph{import} $\mathrm{imp}_{i}^{(S)}(y)$ (second outer circle), \emph{export} $\mathrm{exp}_{i}^{(S)}(y)$ (second inner circle) and \emph{demand} $\mathrm{dem}_{i}^{(S)}(y)$ (inner circle) per country in $y$=1992 (left) and $y$=2013 (right).
   }
 \end{figure}

\begin{figure}[htbp]
 \centering
\includegraphics[width=0.49\textwidth]{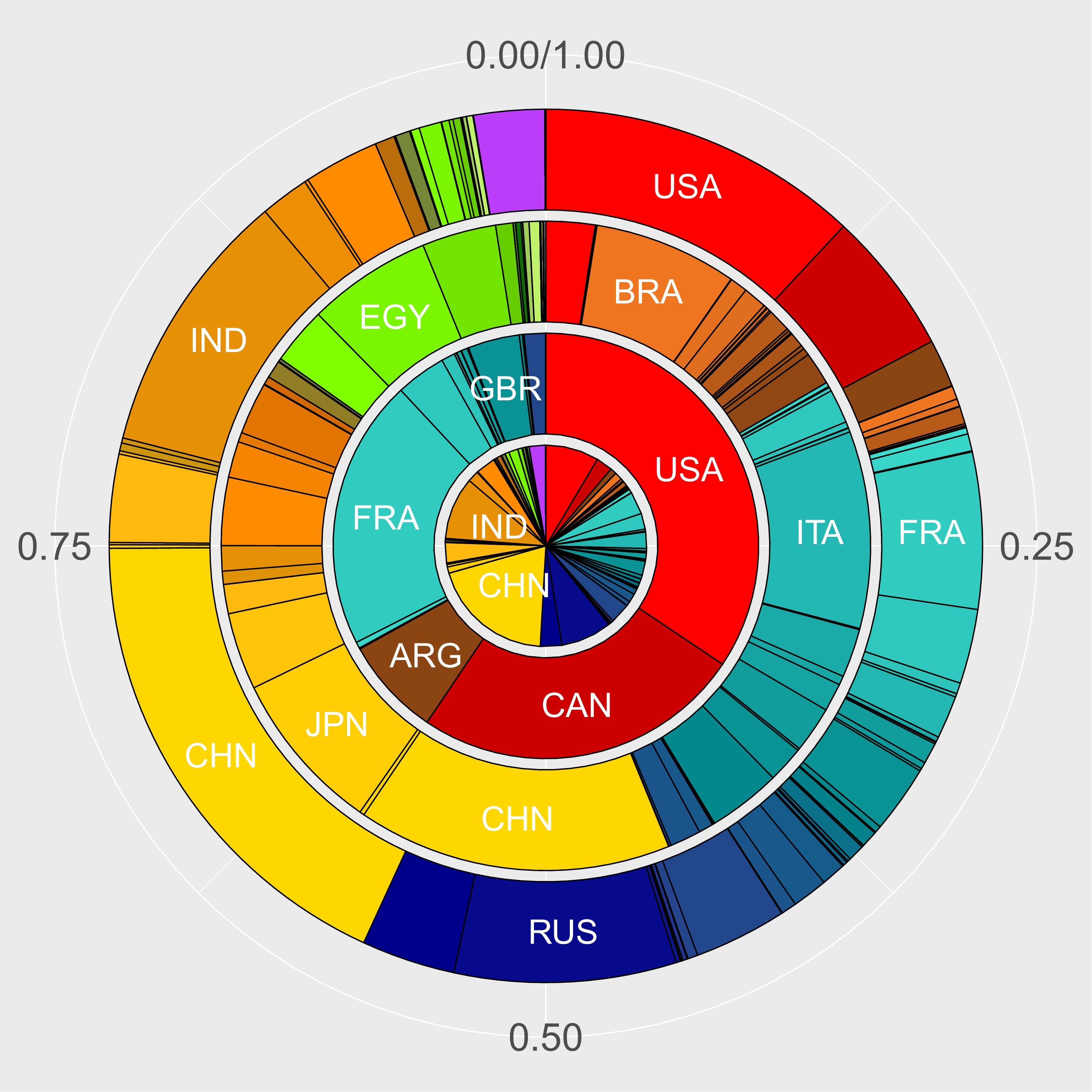}
\includegraphics[width=0.49\textwidth]{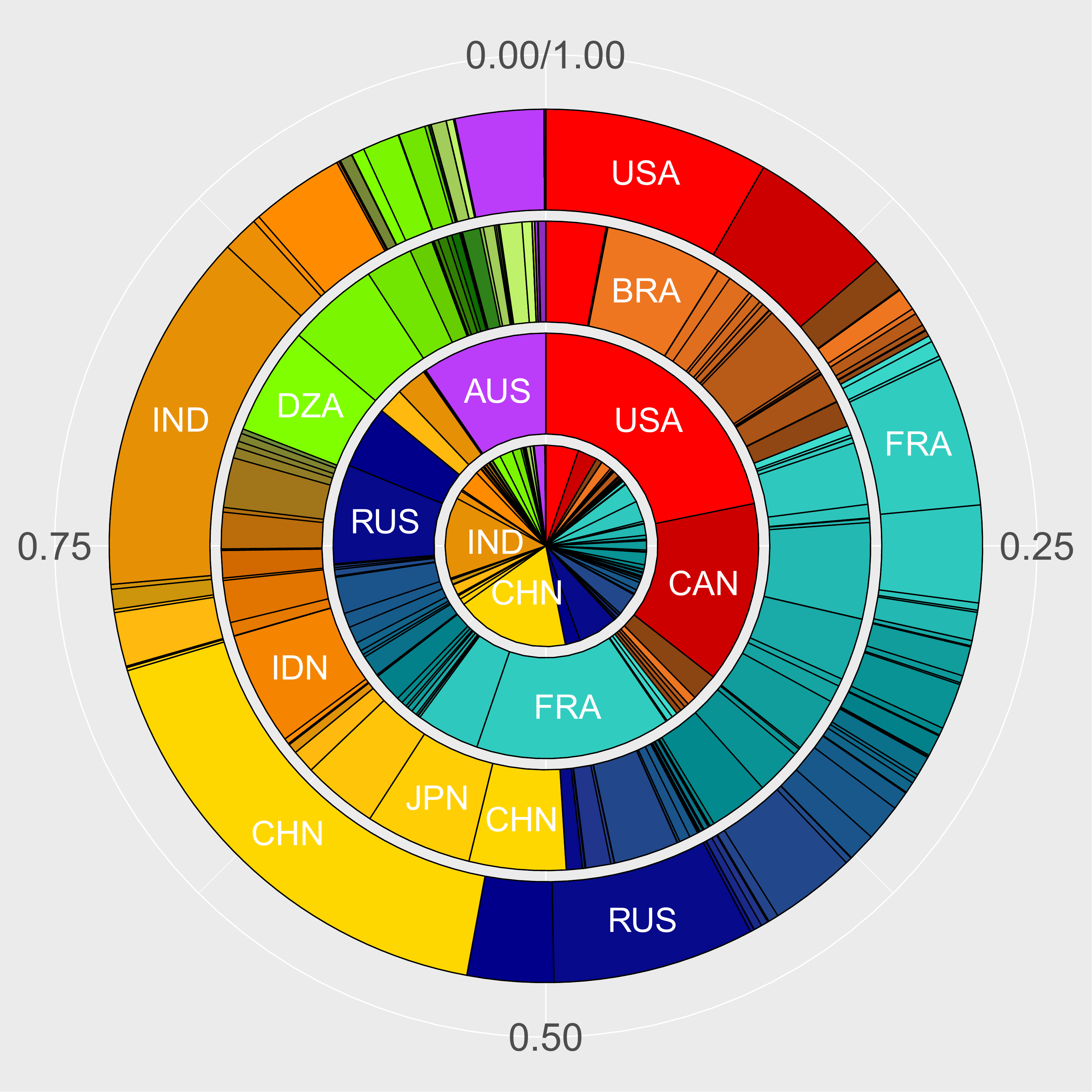}
   \caption{\label{fig:piew} Fractions of wheat \emph{production} $\mathrm{prod}_{i}^{(W)}(y)$ (outer circle), \emph{import} $\mathrm{imp}_{i}^{(W)}(y)$ (second outer circle), \emph{export} $\mathrm{exp}_{i}^{(W)}(y)$ (second inner circle) and \emph{demand} $\mathrm{dem}_{i}^{(W)}(y)$ (inner circle) per country in $y$=1992 (left) and $y$=2013 (right).
   }
 \end{figure}

The Figures \ref{fig:pier}, \ref{fig:pies}, \ref{fig:piew} complement Fig. \ref{fig:pieMaize} for maize. They also refer to the years 1992 (left) and 2013 (right). 

 \clearpage
 
\section{Inequalities in the trade dependency networks} \label{sec:inqu-trade-depend}

\begin{figure}[htbp]
 \subfloat{(a)}{\includegraphics[width=0.41\textwidth]{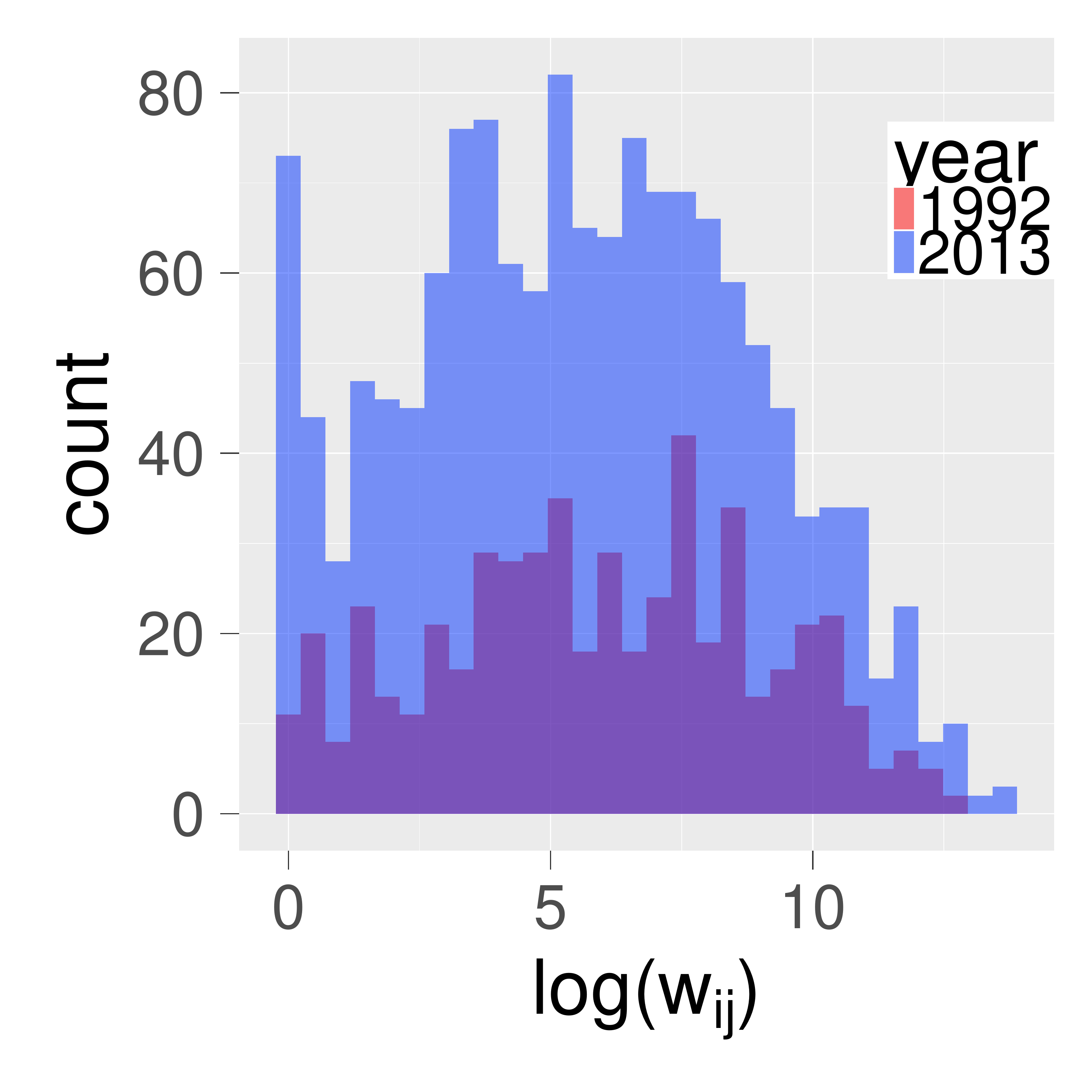}}
\hfill 
 \subfloat{(b)}{\includegraphics[width=0.41\textwidth]{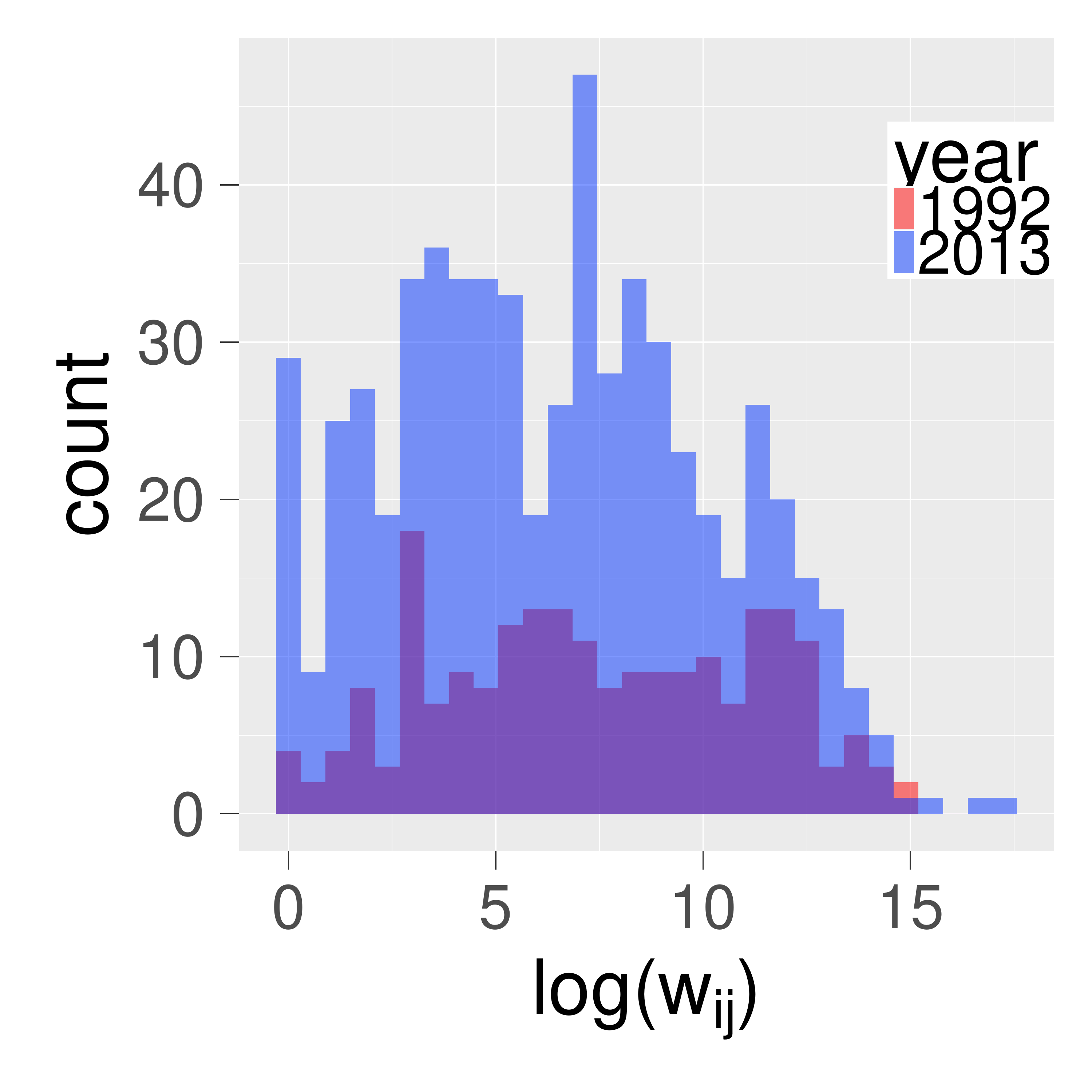}}
 \bigskip
  \subfloat{(c)}{\includegraphics[width=0.41\textwidth]{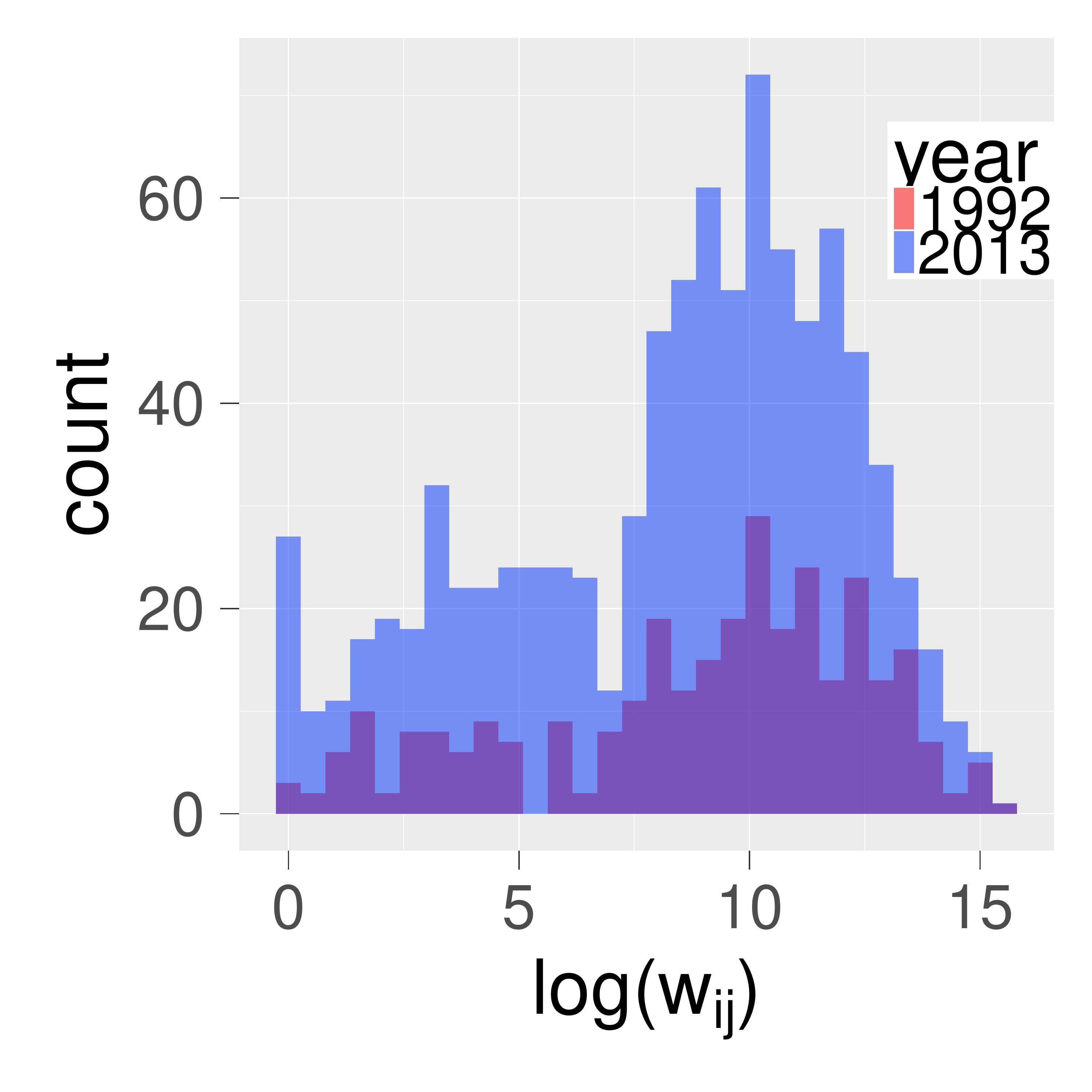}}
\hfill 
 \subfloat{(d)}{\includegraphics[width=0.39\textwidth]{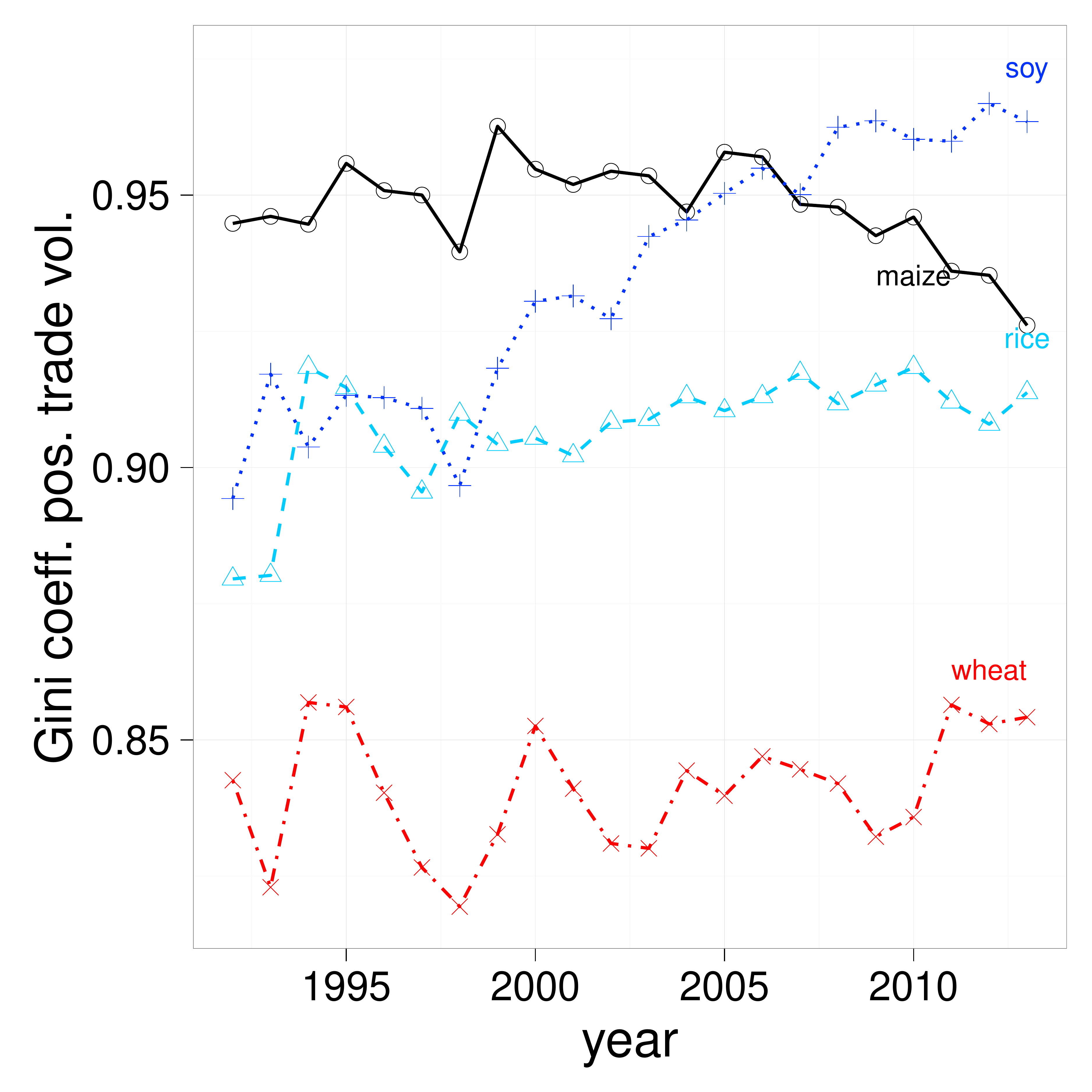}}
 \caption{Histogram of the logarithm of the positive trade volumes in the years 1992 (red or purple if behind the blue) and 2013 (blue) for (a) rice, (b)~soy, (c) wheat. (d) Gini coefficient of the distribution of positive trade volumes over time.}
\label{fig:weights-soy}
\end{figure}

The Figures \ref{fig:weights-soy} (a,b,c) complement Fig.~\ref{fig:weights} for maize. 
Also for rice, soy and wheat we find that the weight distributions are highly right skewed, indicating thatover time.
the trade volumes along a link are very different.
To emphasize this, Fig.~\ref{fig:weights-soy}~(d) shows the evolution of the Gini coefficient \citep{gini1},
which serves as measure for the dissimilarity between positive trade volumes.
We note the rather high values of the Gini coefficient, which do not change much over time.

\clearpage 

\section{Evolution of the global maize trade network}
\label{sec:maize}

\begin{figure}[htbp]
 \centering
 \subfloat{(a)}{\includegraphics[width=0.4\textwidth]{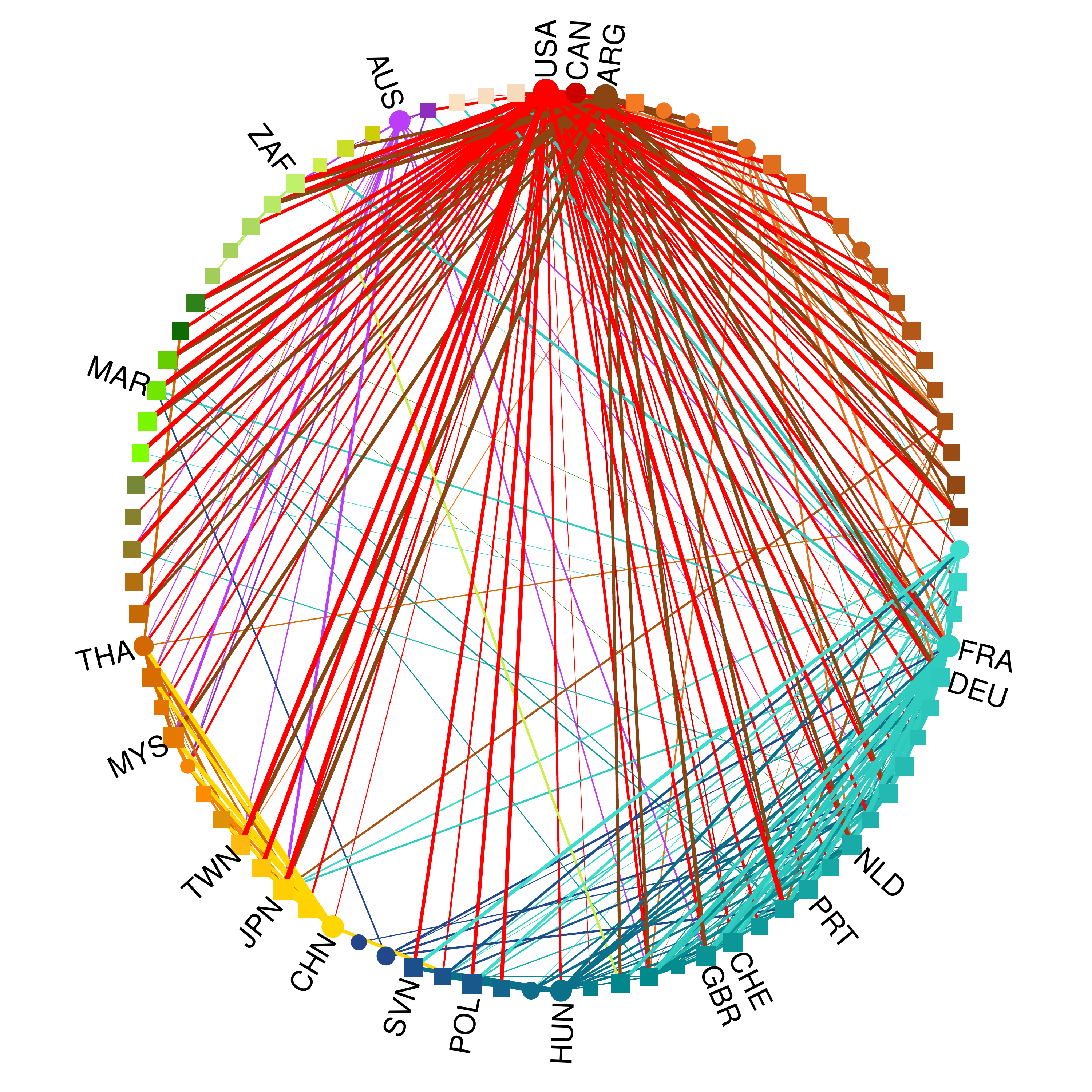}}\subfloat{(b)}{\includegraphics[width=0.4\textwidth]{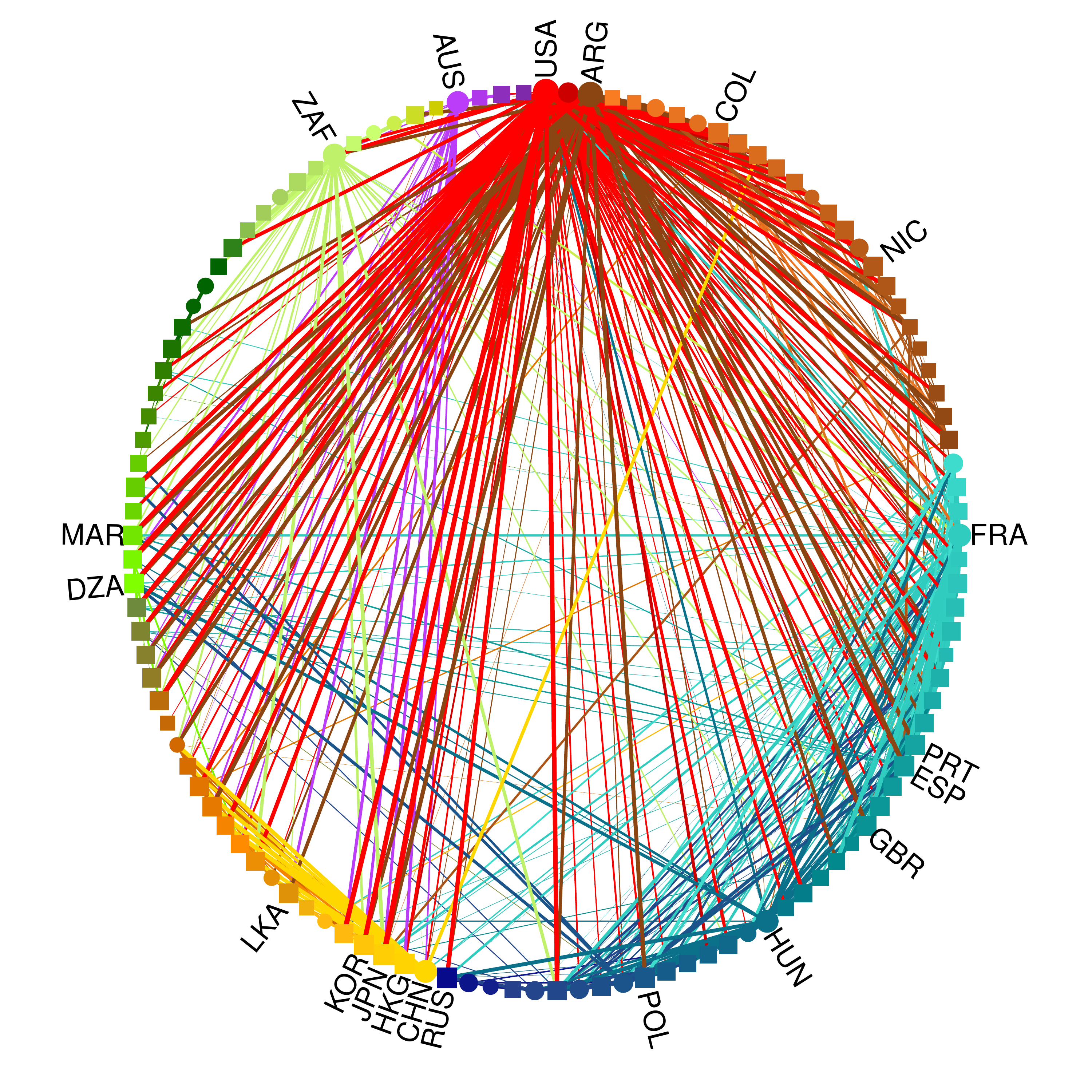}}\\\subfloat{(c)}{\includegraphics[width=0.4\textwidth]{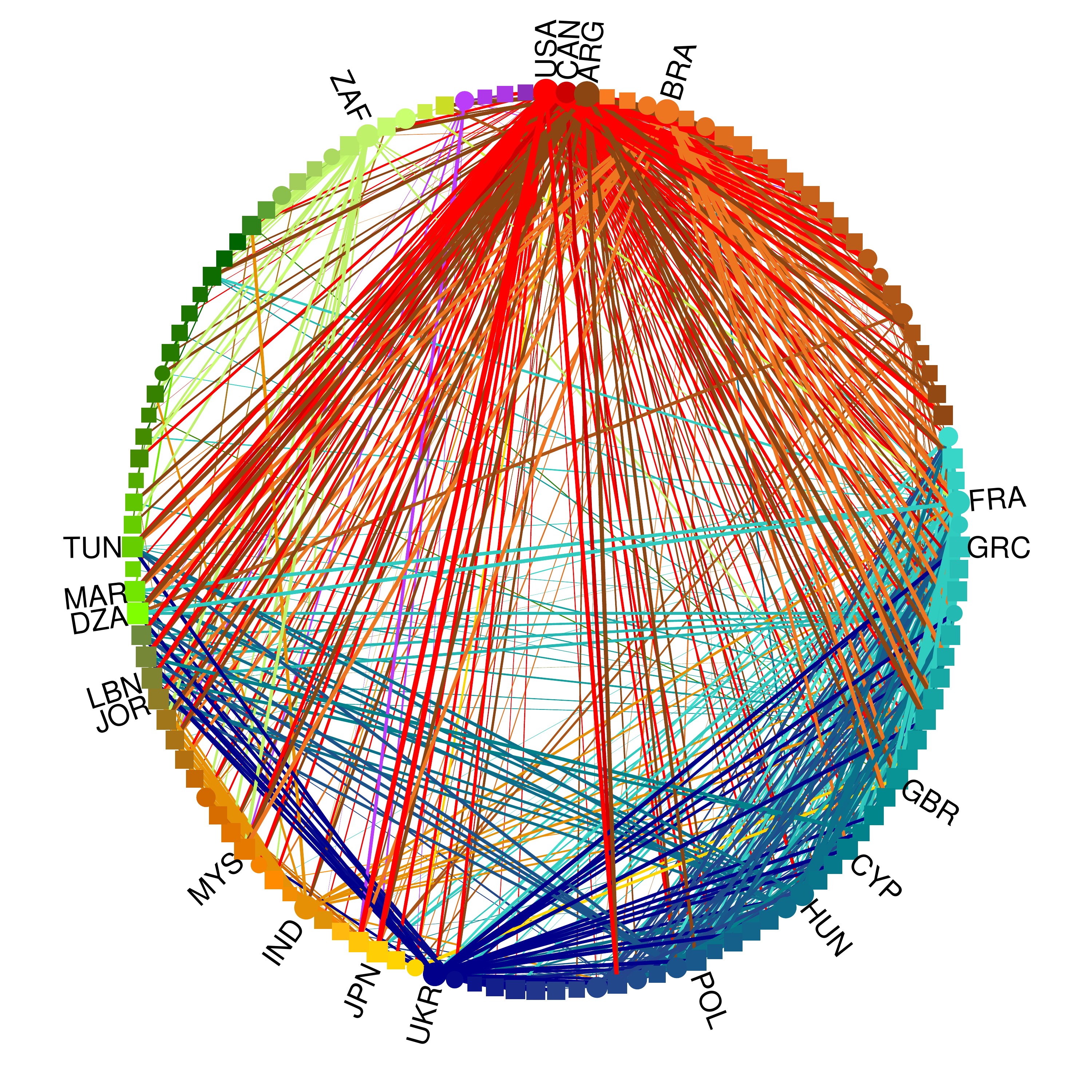}}\subfloat{(d)}{\includegraphics[width=0.4\textwidth]{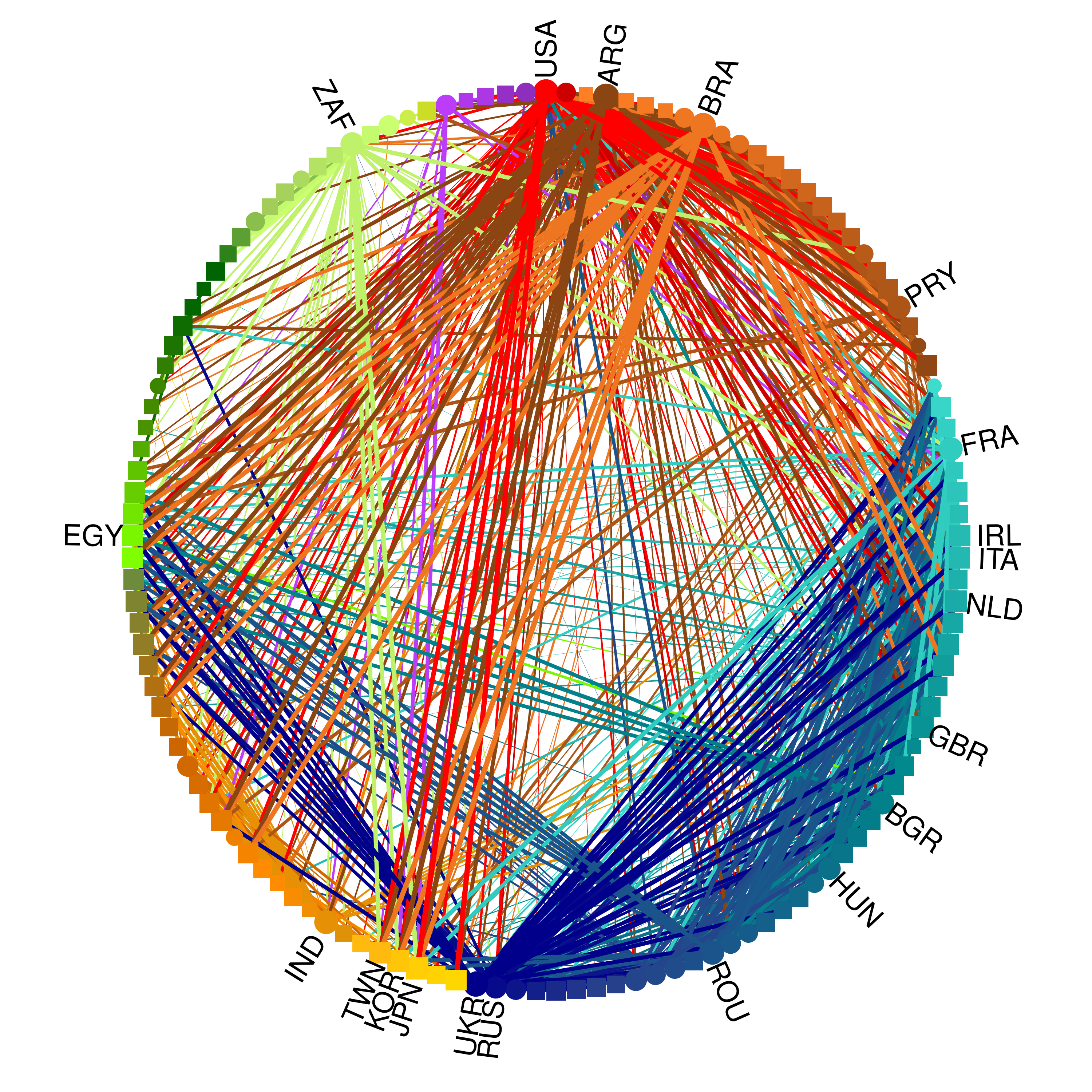}}\caption{Maize trade. Snapshots of the years: (a) 1992, (b) 2000, (c) 2008 and (d) 2013. For further information about the color code, etc., see the caption of Fig.~\ref{fig:data}.
 We observe a clear increase in the number of market participants and in interconnectivity.
In 1992, the USA dominated the international market with almost 46\% of the world wide maize production.
ARG, CAN and AUS are additional bigger exporters that serve other continents, while CHN mainly serves the Asian market, and FRA and HUN export primarily within Europe.
Over the years, further countries cultivate maize so that the share of the production by the USA declines to 35\% in 2013.
Accordingly, additional exporters and importers enter the market.
}
\label{fig:wMnet}
\end{figure}

\clearpage

\section{Evolution of the global rice trade network}
\label{sec:furth-avail-rice}

\begin{figure}[htbp]
 \centering
 \subfloat{(a)}{\includegraphics[width=0.4\textwidth]{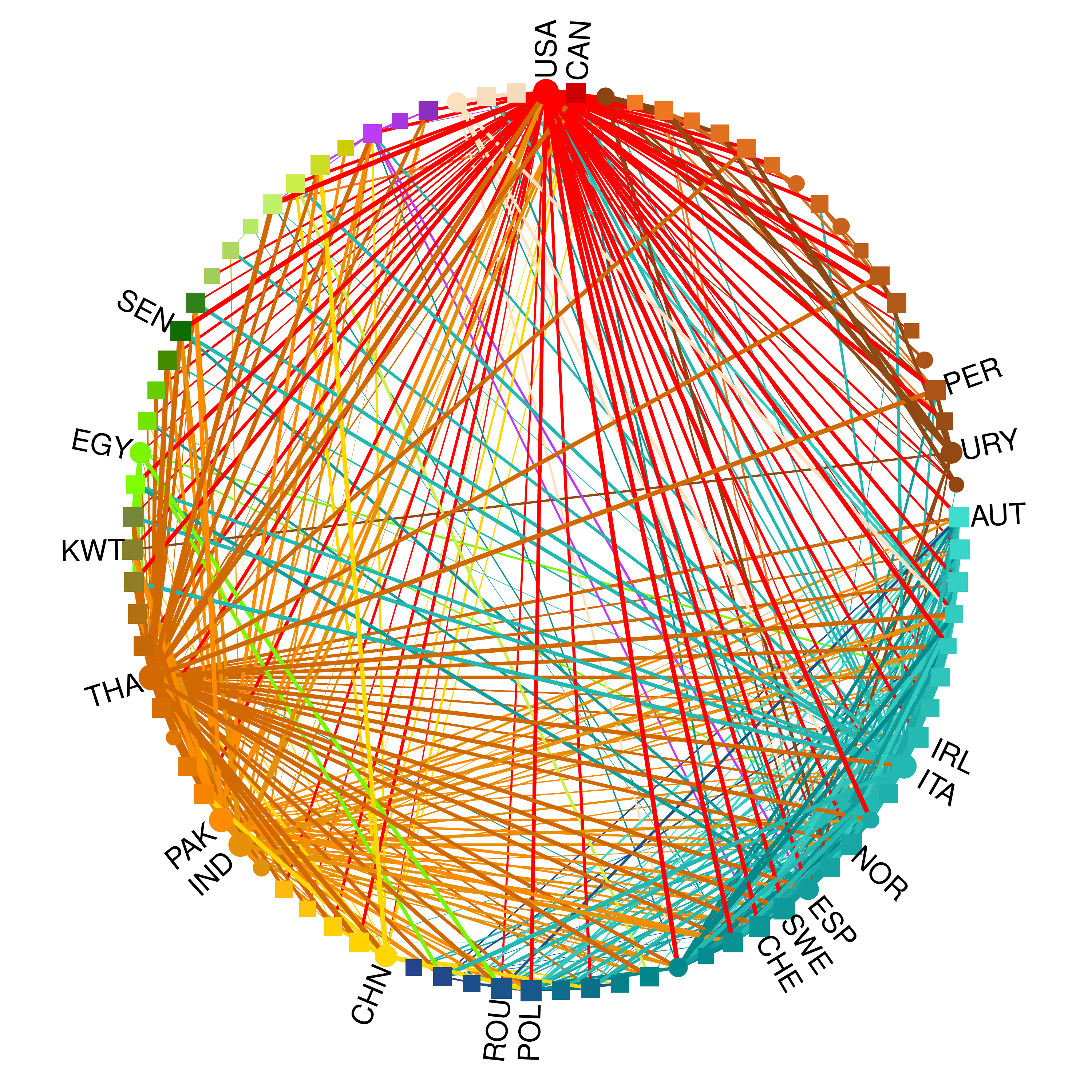}}\subfloat{(b)}{\includegraphics[width=0.4\textwidth]{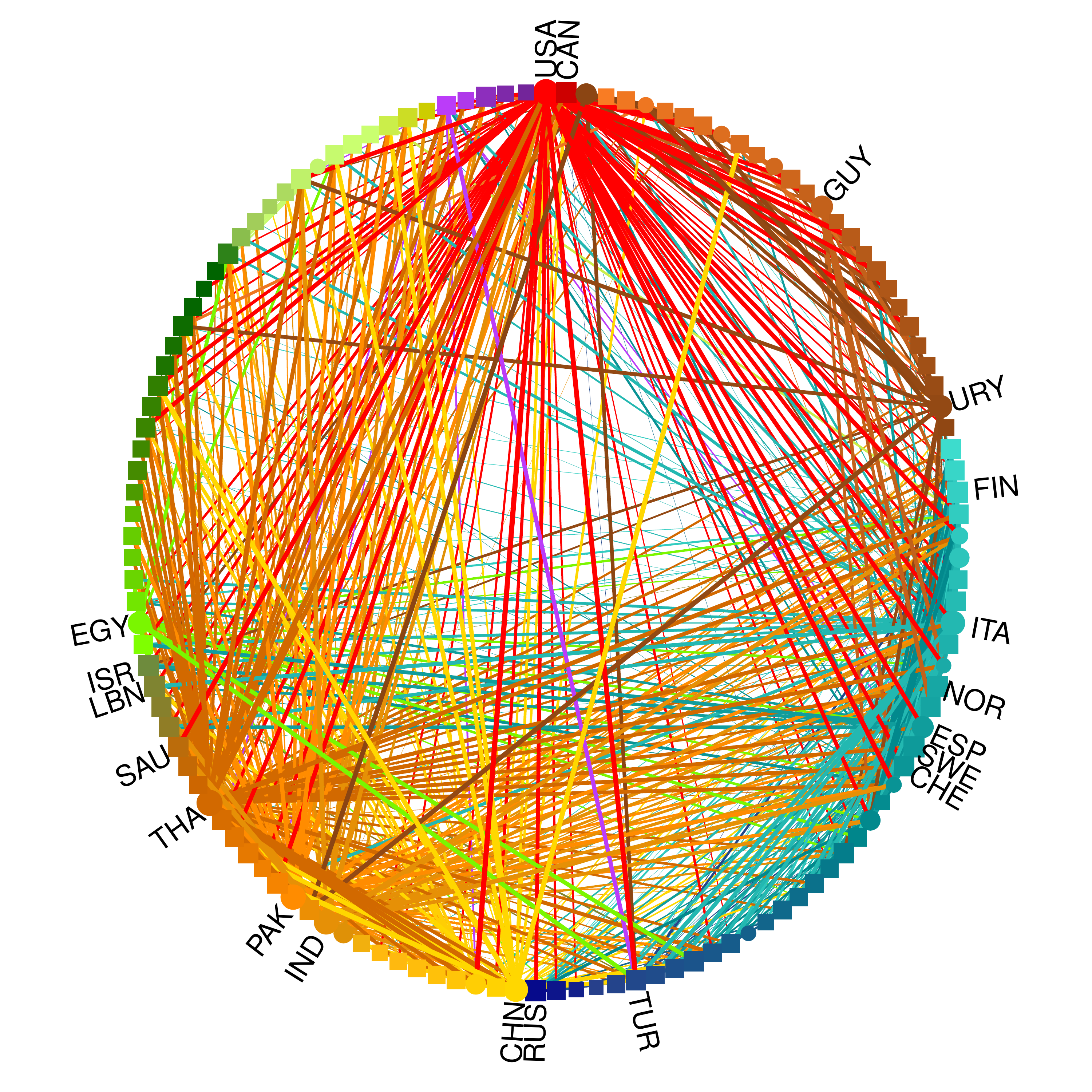}}\\\subfloat{(c)}{\includegraphics[width=0.4\textwidth]{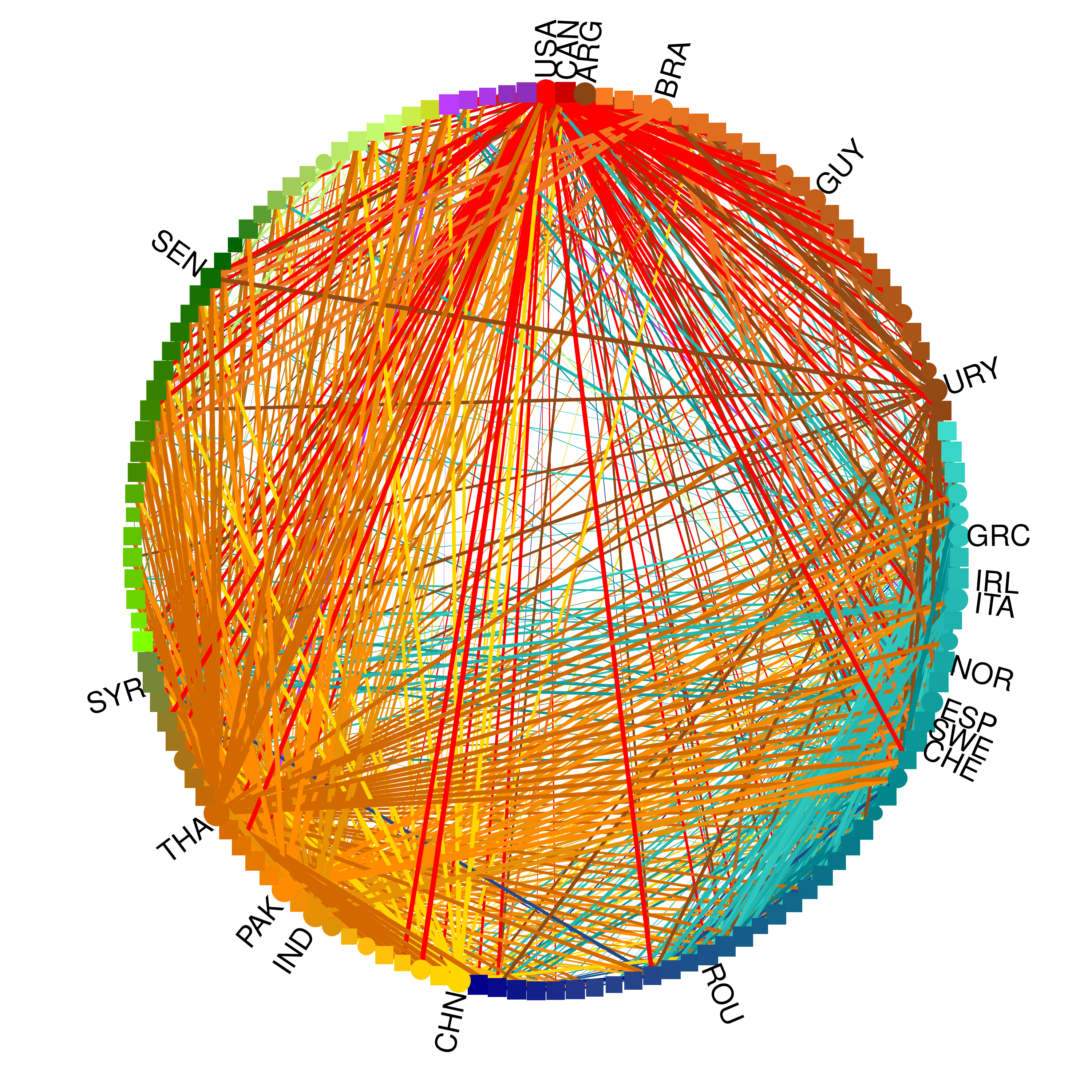}}\subfloat{(d)}{\includegraphics[width=0.4\textwidth]{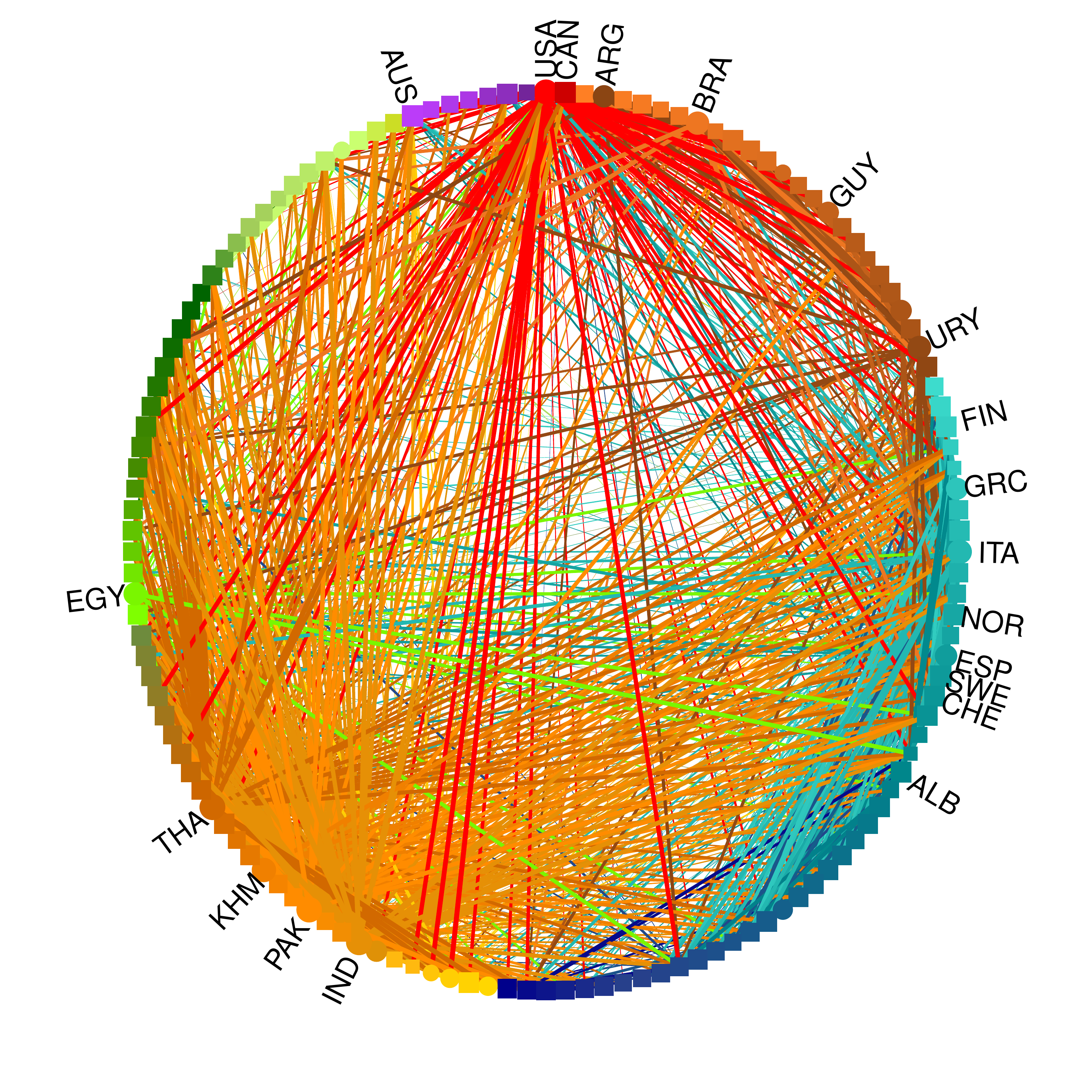}}\caption{Rice trade. Evolution of the international rice trade network. Snapshots of the years: (a) 1992, (b) 2000, (c) 2008 and (d) 2013. For further information about the color code, etc., see the caption of Fig.~\ref{fig:data}.
  As for maize, we observe an increasing interconnectivity  in the global trade of rice over the years.
Still, only about 4\% of the total production is traded in 2013 because 
national production is often subject to export (or even import) restrictions or other protective policies.
In Asia, rice is primarily produced for national consumption, while overproduction is traded.
}
\label{fig:wRnet}
\end{figure}

\clearpage

\section{Evolution of the global soy trade network}
\label{sec:furth-avail-soy}

\begin{figure}[htbp]
 \centering
 \subfloat{(a)}{\includegraphics[width=0.4\textwidth]{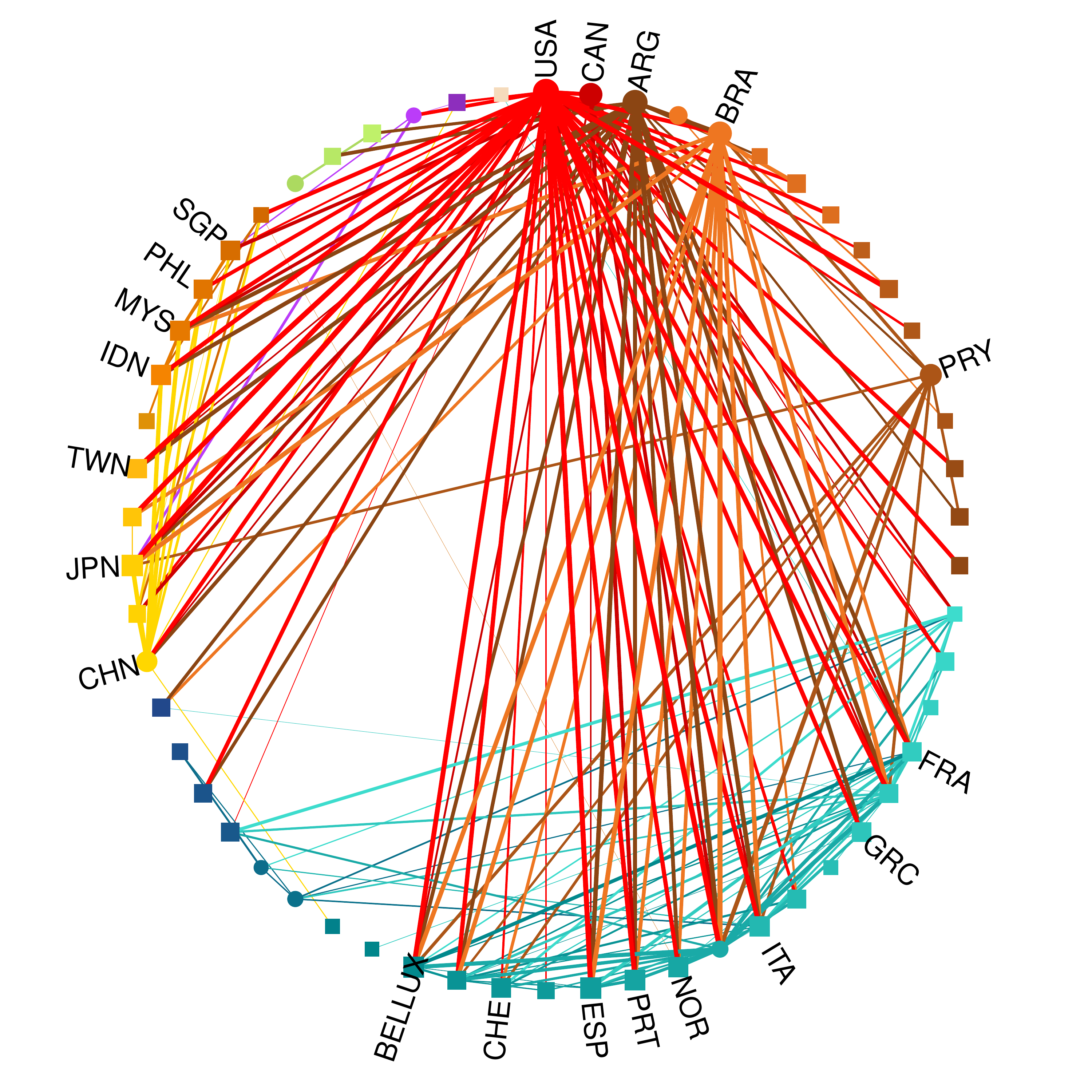}}\subfloat{(b)}{\includegraphics[width=0.4\textwidth]{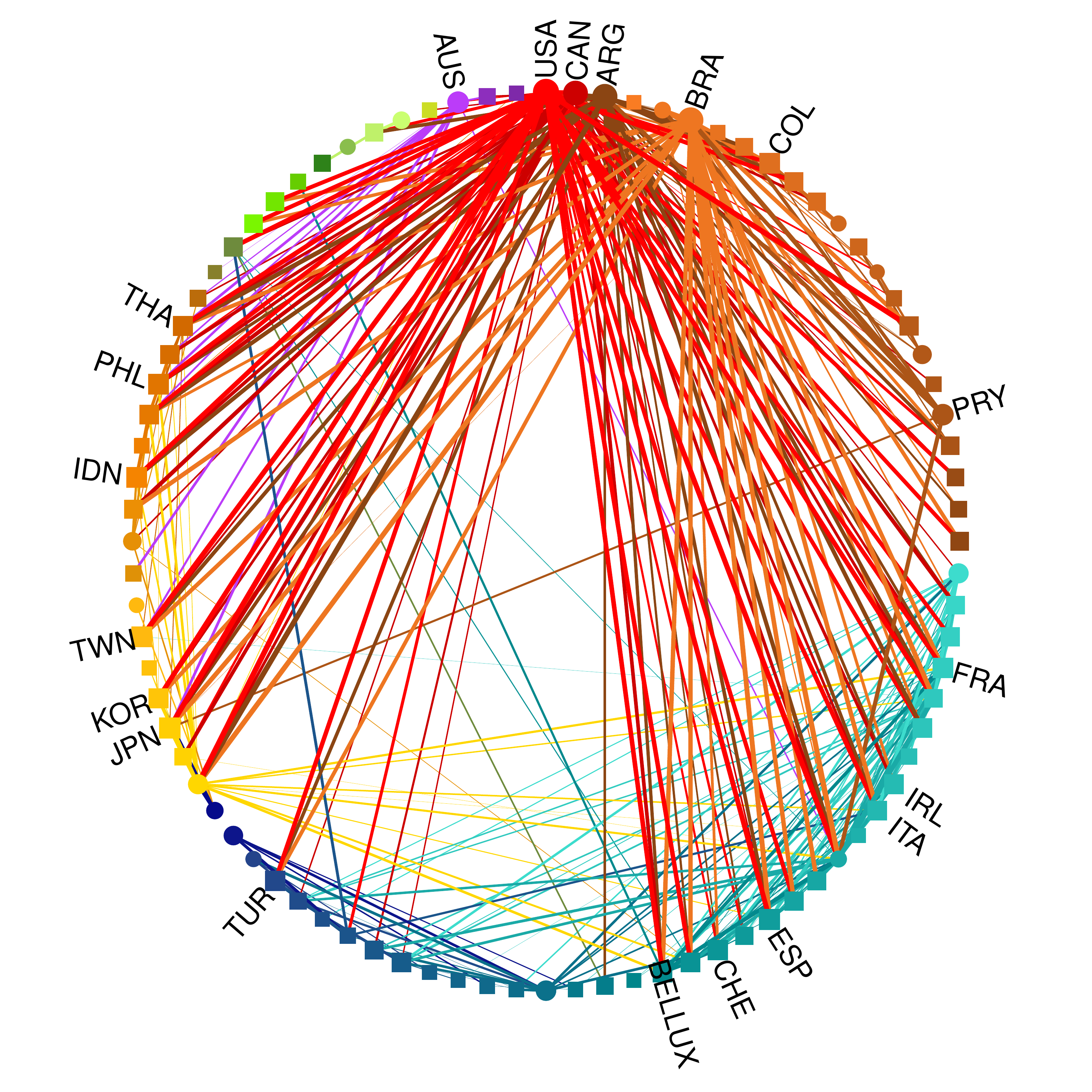}}\\\subfloat{(c)}{\includegraphics[width=0.4\textwidth]{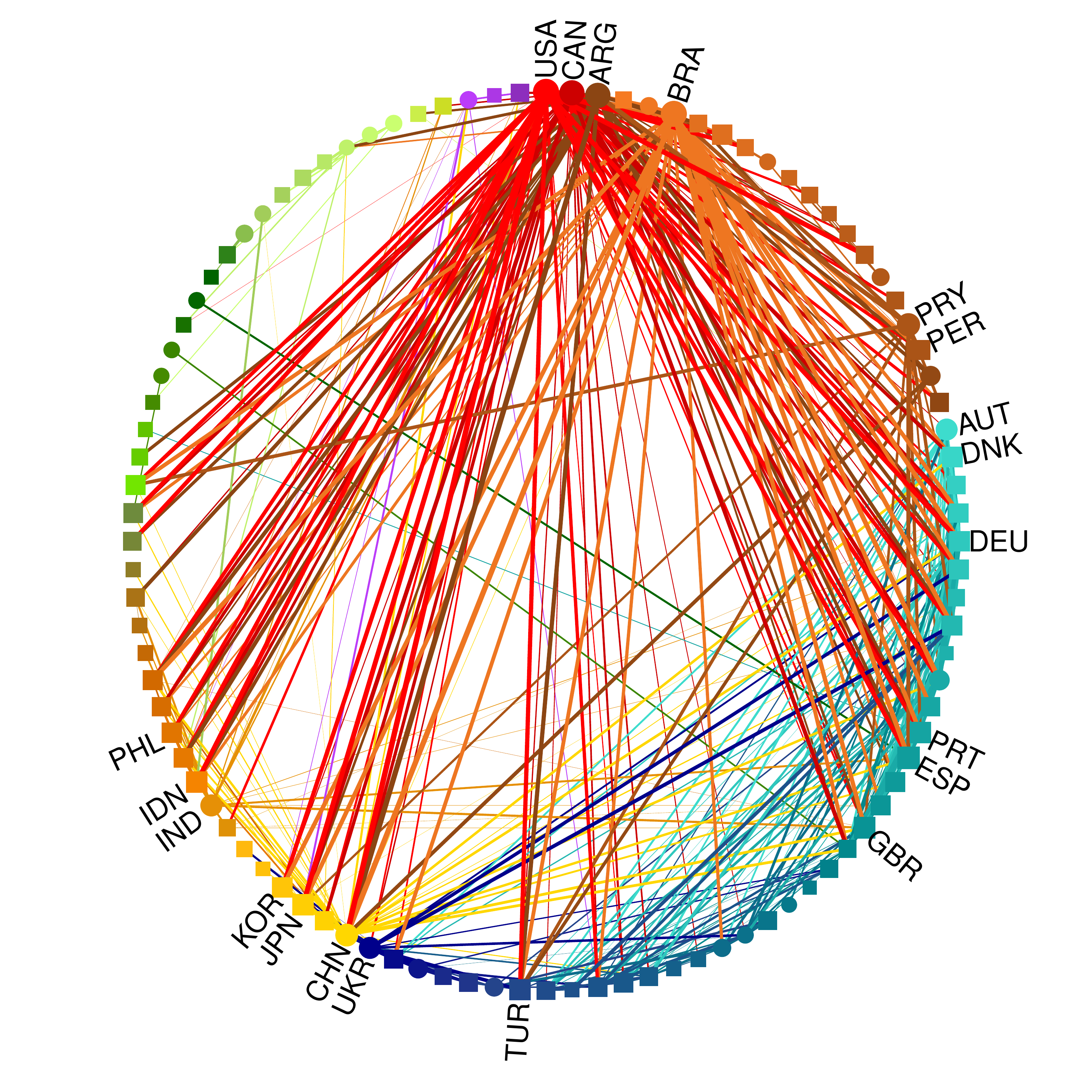}}\subfloat{(d)}{\includegraphics[width=0.4\textwidth]{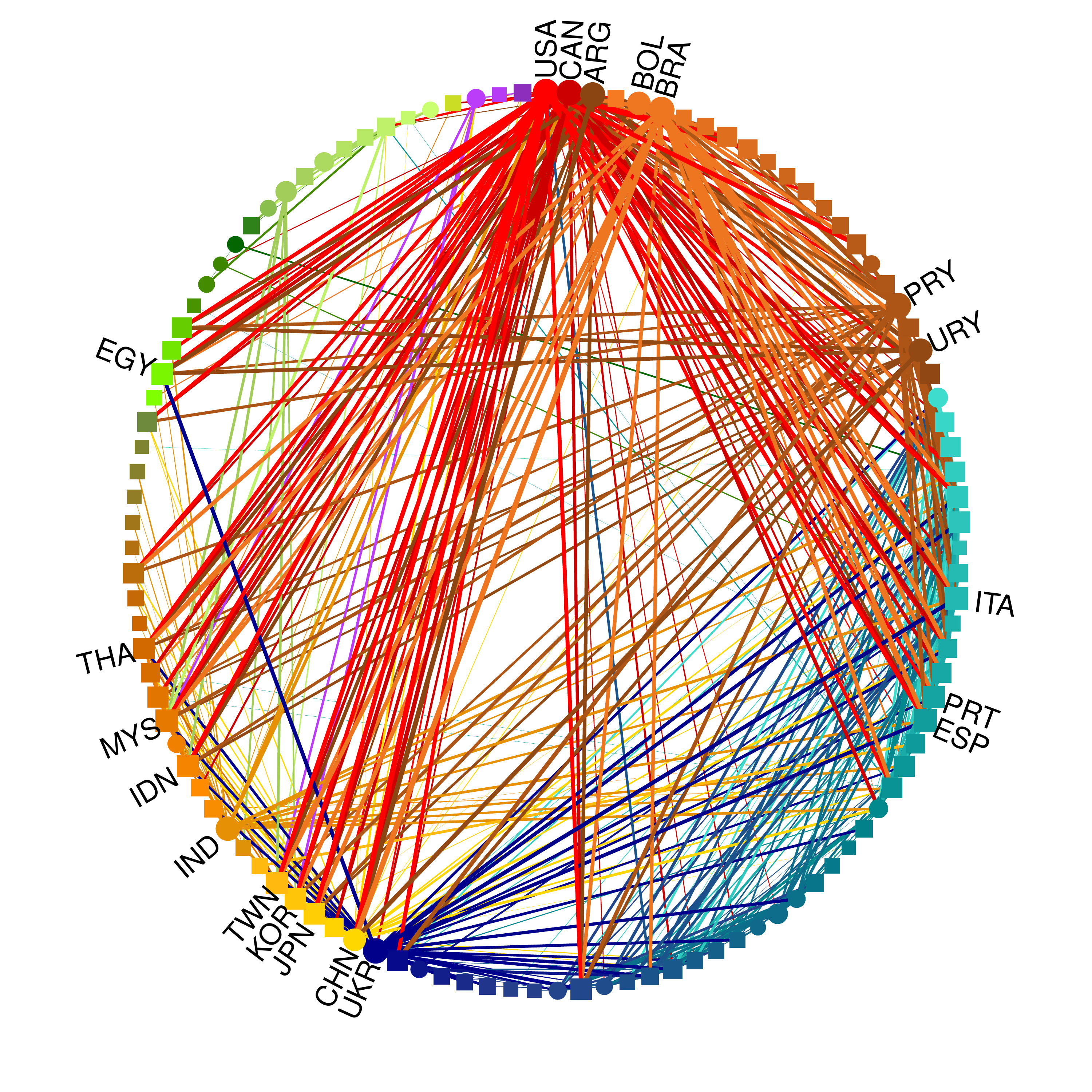}}\caption{Soybean trade. Snapshots of the years: (a) 1992, (b) 2000, (c) 2008 and (d) 2013. For further information about the color code, etc., see the caption of Fig.~\ref{fig:data}.
 This global trade network seems to be less dense than the ones for maize, rice, or wheat, although we observe growing interconnectivity.
With an increasing number of market participants, also a diversification of production and trade links is associated.
While in 1992 the USA produces 52\% of the total production, this share decreases to 33\% in 2013 and BRA produces a similar amount of soy.
Whereas CHN is a net exporter in 1992, it imports far more than it produces in 2013.
}
\label{fig:wSnet}
\end{figure}

\clearpage 

\section{Evolution of the global wheat trade network}
\label{sec:furth-avail-wheat}

\begin{figure}[htbp]
 \centering
\includegraphics[width=0.4\textwidth]{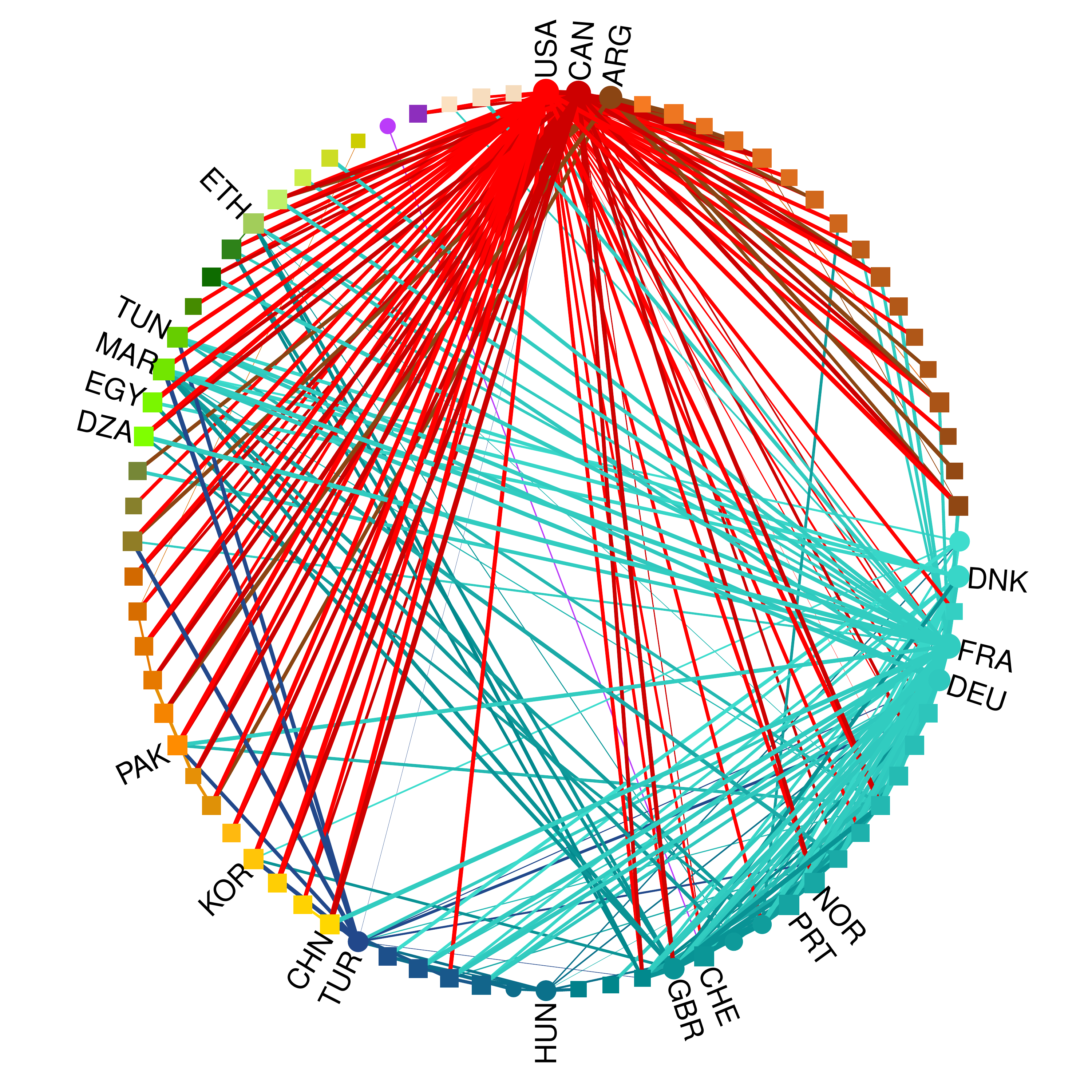}\raisebox{3ex}{(a)}
\includegraphics[width=0.4\textwidth]{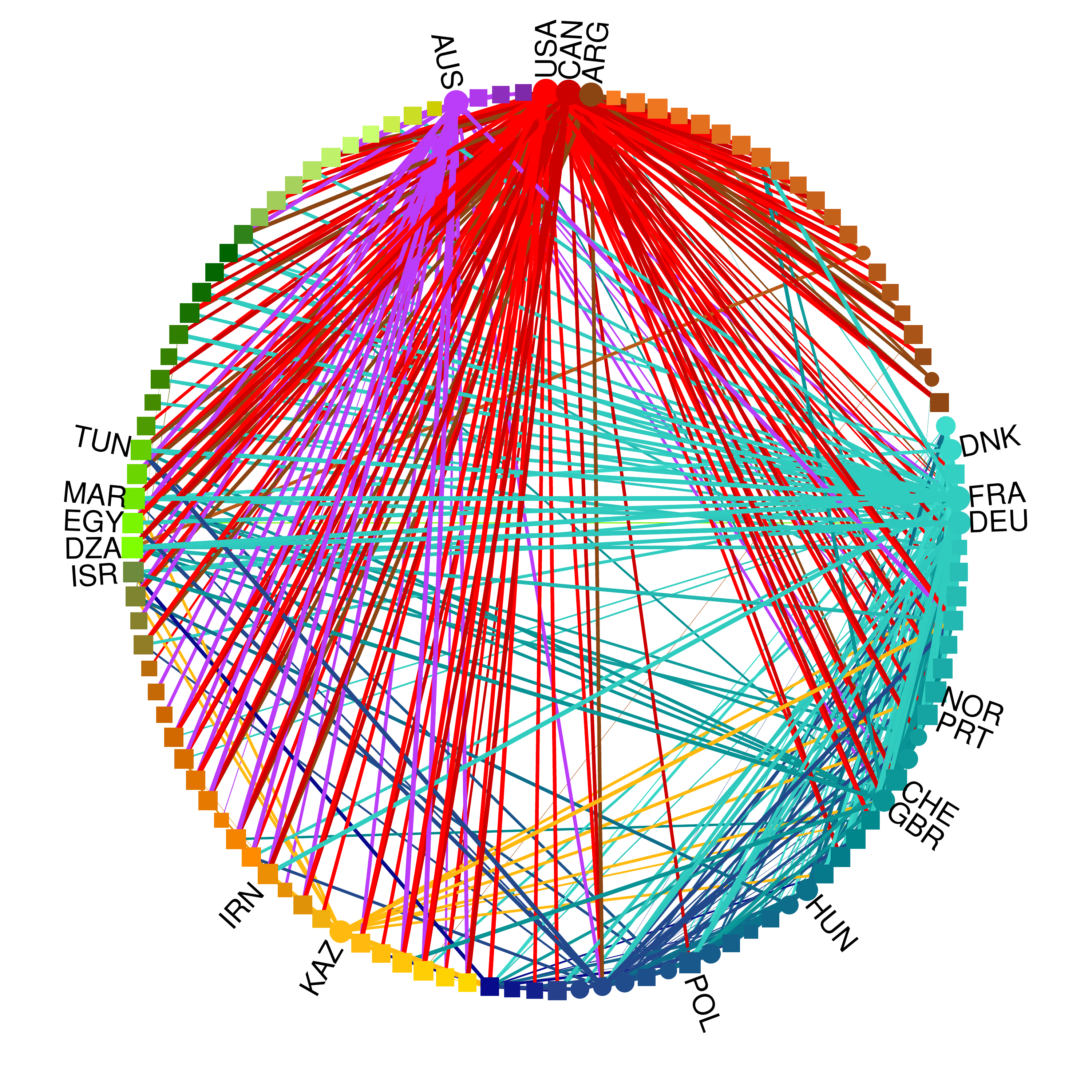}\raisebox{3ex}{(b)}\\
\includegraphics[width=0.4\textwidth]{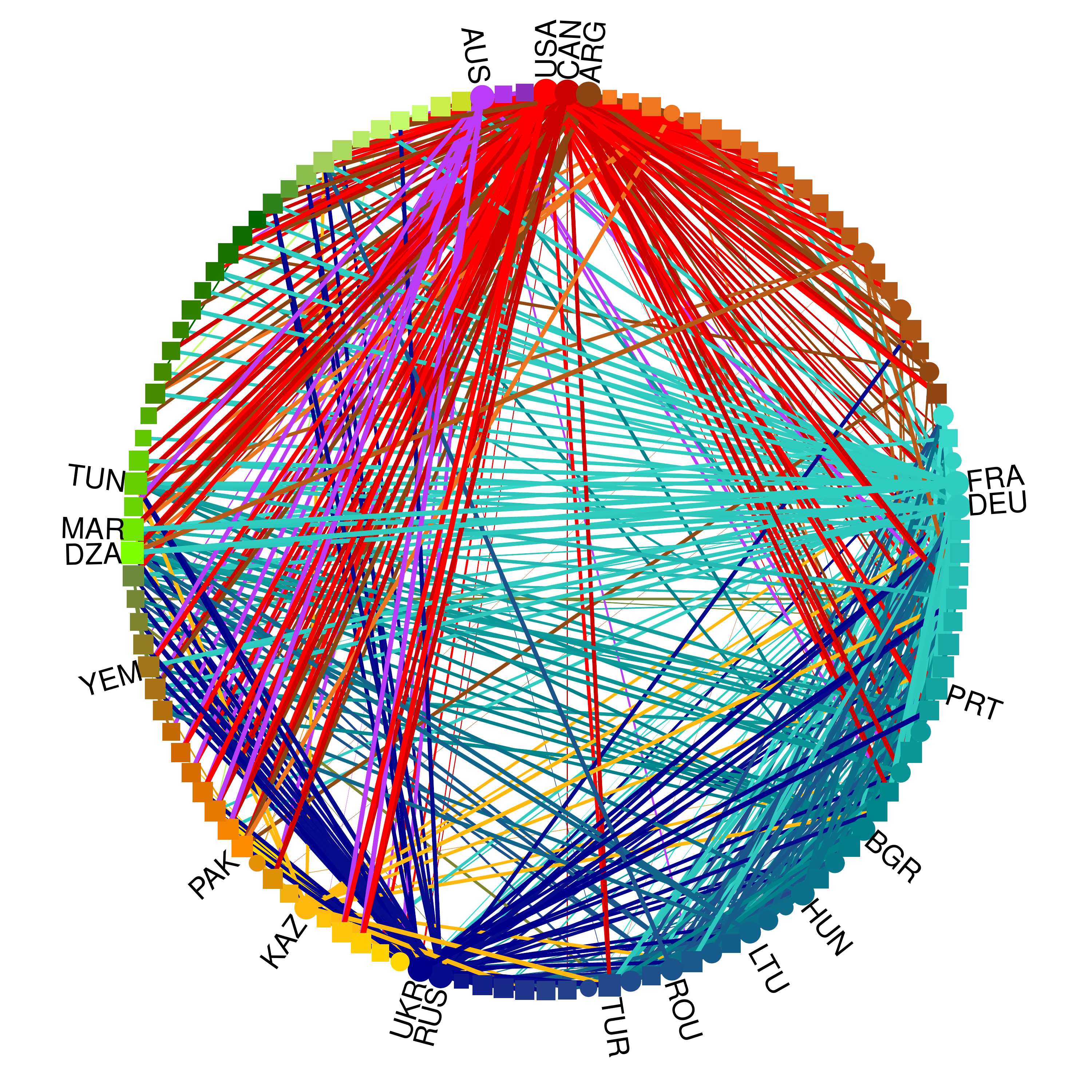}\raisebox{3ex}{(c)}
\includegraphics[width=0.4\textwidth]{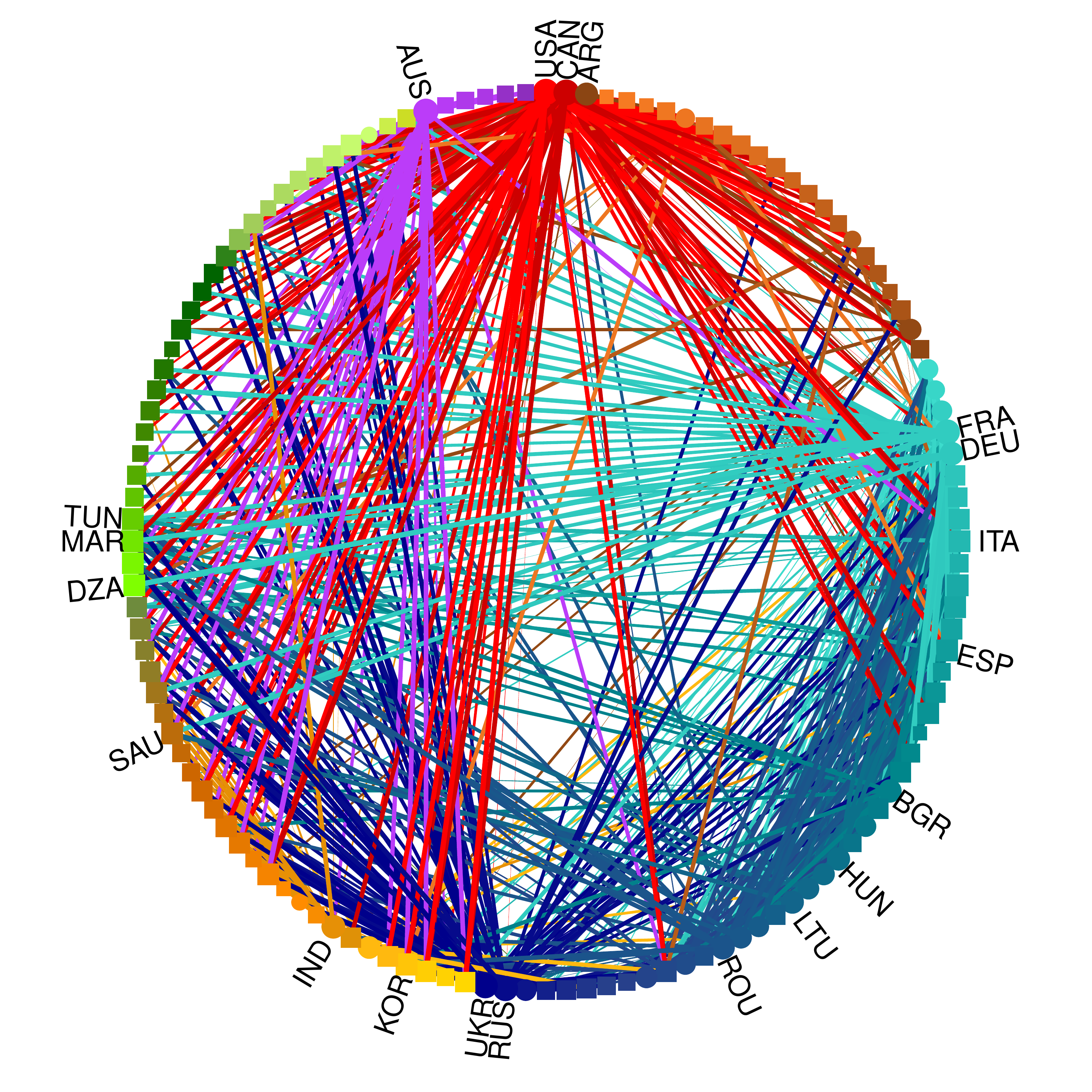}\raisebox{3ex}{(d)}
\caption{Wheat trade. Snapshots of the years: (a) 1992, (b) 2000, (c) 2008 and (d) 2013. For further information about the color code, etc., see the caption of Fig.~\ref{fig:data}.
 The USA, CAN, and ARG, as well as several European countries, for instance, FRA, DEU, and HUN, together with AUS are the main exporting nations.
Especially since 2000, UKR and RUS as big producers and exporters add  several links to the network and increase interconnectivity.
}
\label{fig:wWnet}
\end{figure}

\end{appendix}

\end{document}